\documentclass[longauth]{aa}
\usepackage{graphicx}
\usepackage{txfonts}
\usepackage{hyperref}

\usepackage[utf8]{inputenc}
\usepackage[T1]{fontenc}
\usepackage{amsfonts}

\usepackage{float}
\usepackage{subcaption}
\usepackage{graphicx}

\usepackage{makecell}

\usepackage{pgfplots}
\pgfplotsset{compat=1.16}

\newcommand{\emr}[1]{\ensuremath{\mathrm{#1}}}
\newcommand{\emm}[1]{\ensuremath{#1}}
\newcommand{\e}[1]{\emm{\times 10^{#1}}}
\newcommand{\ch}[1]{\ensuremath{\mathrm{#1}}}
\newcommand{\unit}[1]{\ensuremath{\,\mathrm{#1}}}

\newcommand{\thCO}{\ch{^{13}CO}}
\newcommand{\twCO}{\ch{^{12}CO}} 
\newcommand{\CeiO}{\ch{C^{18}O}}
\newcommand{\CseO}{\ch{C^{17}O}}

\newcommand{\Jone}{\ch{(1-0)}}
\newcommand{\J}[2]{$J$=#1$-$#2}

\newcommand{\pc}{\unit{pc}} 
\newcommand{\mpc}{\unit{mpc}} 
\newcommand{\kms}{\unit{km\,s^{-1}}}
\newcommand{\kHz}{\unit{kHz}}

\newcommand{\GHz}{\unit{GHz}} 

\newcommand{\radec}[6]{\emr{#1^{h}#2^{m}#3^{s},#4^{\circ}#5^{'}#6^{''}}}

\newcommand{\mi}{\mathcal{I}\xspace}

\newcommand{\paren}[1]  {\left(  #1 \right)}  
\newcommand{\cbrace}[1] {\left\{ #1 \right\}} 
\newcommand{\bracket}[1]{\left[  #1 \right]}  
\newcommand{\modulus}[1]{\left|  #1 \right|}  
\newcommand{\norm}[2]{\left|\left| #2 \right|\right|_{#1}} 

\newcommand{\Esp}[1]{\mathbb{E}\left[#1\right]}
\newcommand{\ft}[1]{\mathcal{F}\bracket{#1}}

\newcommand{\sinC}[1]{\emm{\mathrm{sinC}\left(#1\right)}}

\newcommand{\ESD}{\emm{\mathcal{E}}\xspace}
\newcommand{\PSD}{\emm{\mathcal{P}}\xspace}

\newcommand{\ie}{\textit{i.e.}\xspace}
\newcommand{\eg}{\textit{e.g.}\xspace}

\newcommand{\Int}[1]{\ensuremath{I_{\text{#1}}}}
\newcommand{\Sig}[1]{\ensuremath{S_{\text{#1}}}}
\newcommand{\Noi}[1]{\ensuremath{N_{\text{#1}}}}

\newcommand{\Area}[1]{\emm{A_{\text{#1}}}}
\newcommand{\numb}[1]{\emm{n_{\text{#1}}}}

\DeclareFontFamily{U}{wncy}{}
\DeclareFontShape{U}{wncy}{m}{n}{<->wncyr10}{}
\DeclareSymbolFont{mcy}{U}{wncy}{m}{n}
\DeclareMathSymbol{\Sha}{\mathord}{mcy}{"58} 

\newcommand{\TabDataLines}{%
  \begin{table}
    \centering %
    \caption{Studied molecular lines}
    \begin{tabular}{ccc}
      \hline
      \hline
      Species & Transition & Rest frequency \\
              &            & [GHz]          \\
      \hline
      \thCO{} & \J10 & 110.201354 \\
      \CseO{} & \J10 & 112.358982 \\
      \hline
    \end{tabular}
    \label{tab:lines}
  \end{table}
}

\newcommand{\FigDataImages}{%
  \begin{figure}
    \centering %
    \includegraphics[width=\linewidth]{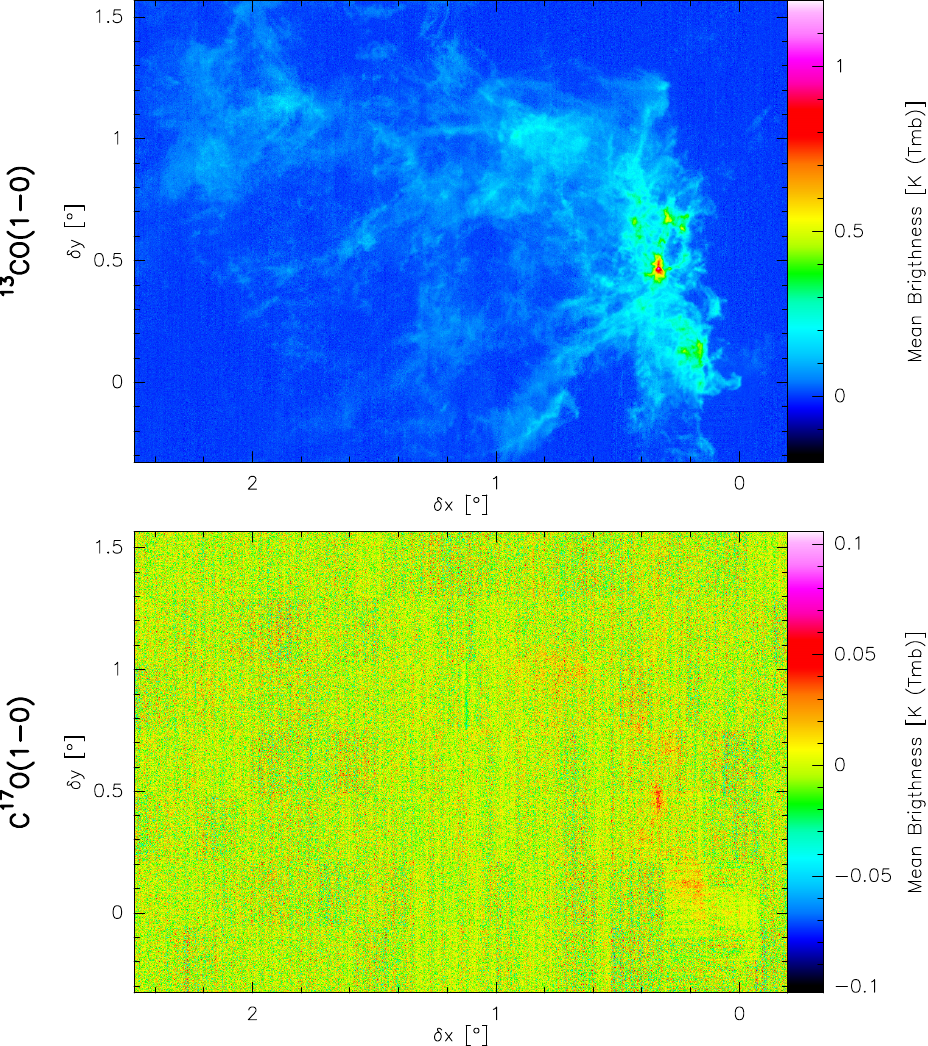}
    \caption{Comparison of mean intensity images between two radio-astronomy lines.}
    \label{fig:data:images}
  \end{figure}
}

\newcommand{\FigDataSpectra}{%
  \begin{figure}
    \centering %
    \includegraphics[width=\linewidth]{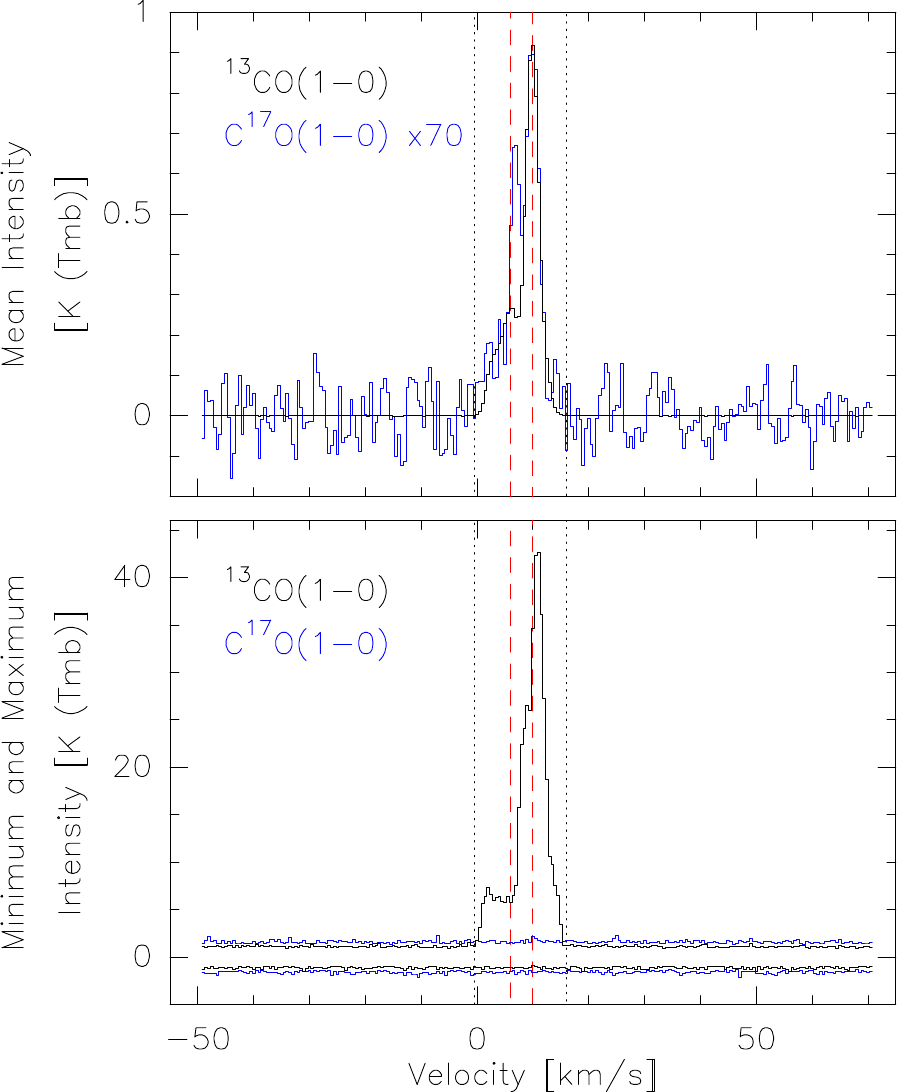}
    \caption{Comparison of intensity spectra between the two
      radio-astronomy lines. The spectra show the mean \textbf{(top)},
      minimum and maximum \textbf{(bottom)} intensity as a function of the
      channel velocity or number. The vertical dashed red lines show the
      channels whose spatial distribution is plotted on
      Fig.~\ref{fig:data:selected:channels}. The vertical dotted lines on
      the radio-astronomy spectra separate the signal channels from the
      noise-only ones.}
    \label{fig:data:spectra}
  \end{figure}
}

\newcommand{\FigDataHisto}{%
  \begin{figure}
    \centering %
    \includegraphics[width=\linewidth]{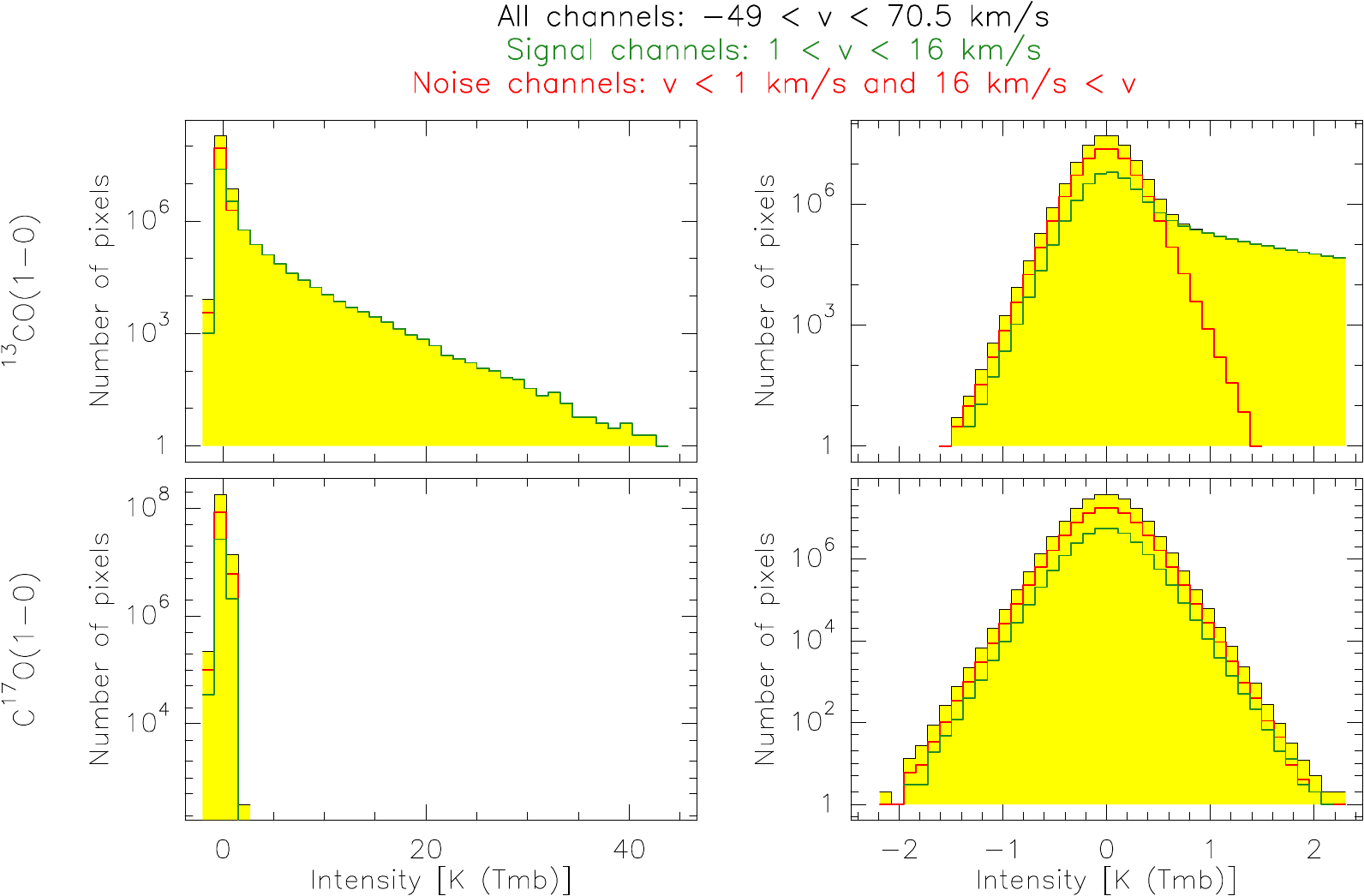}
    \caption{Comparison of the histograms of the intensity of the two
      radio-astronomy lines. The left column shows the full intensity
      dynamical range, while the right column zoom on faint intensity. The
      black histogram is computed over all the data channels. The red and
      green histograms are computed over the channel ranges that contains
      either mostly noise or high signal-to-noise ratio intensity,
      respectively.}
    \label{fig:data:histo}
  \end{figure}
}

\newcommand{\FigDataNoiseDistribution}{%
  \begin{figure}
    \centering %
    \includegraphics[width=\linewidth]{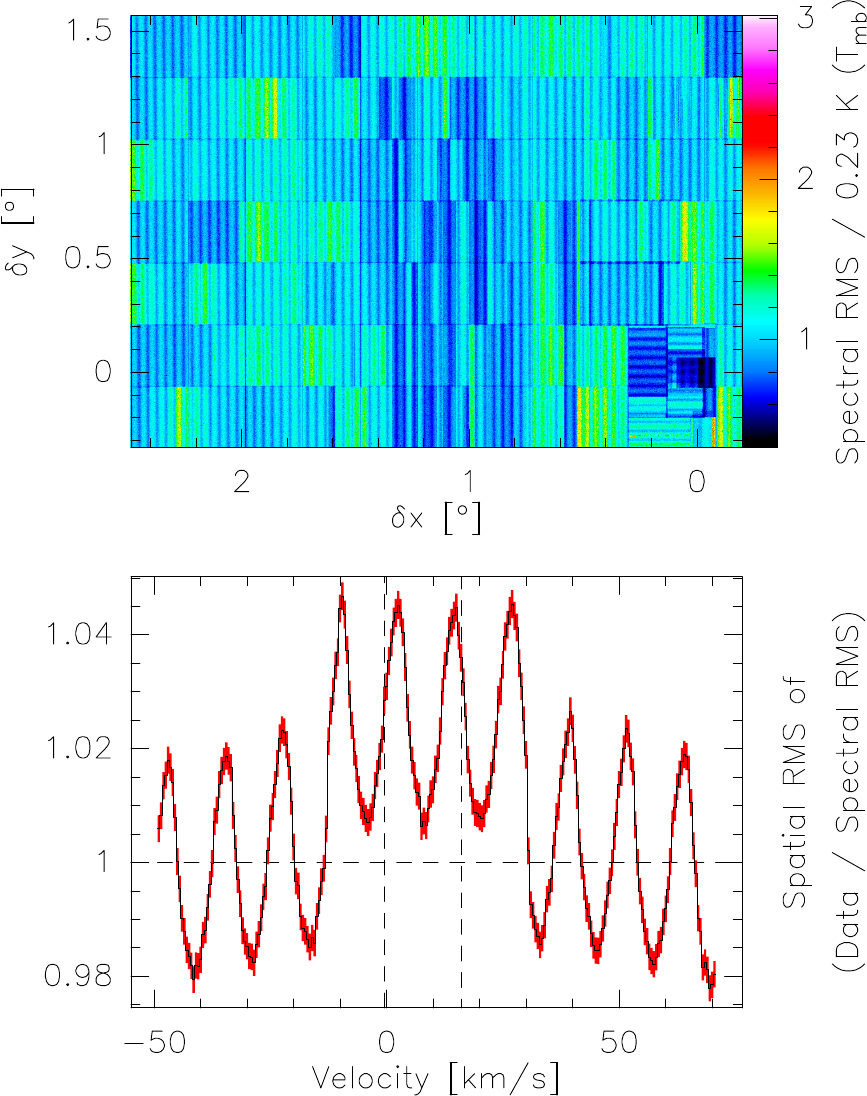}
    \caption{Noise spatial \textbf{(top)} and spectral \textbf{(bottom)}
      variations for the \CseO{} \Jone{} line cube.  The spatial maps were
      normalized by the median noise value. The red region in the bottom
      panels shows the $3\sigma$ uncertainty interval of the computation.}
    \label{fig:data:noise:dist}
  \end{figure}
}

\newcommand{\FigDataSelectedChannels}{%
  \begin{figure}
    \centering %
    \includegraphics[width=\linewidth]{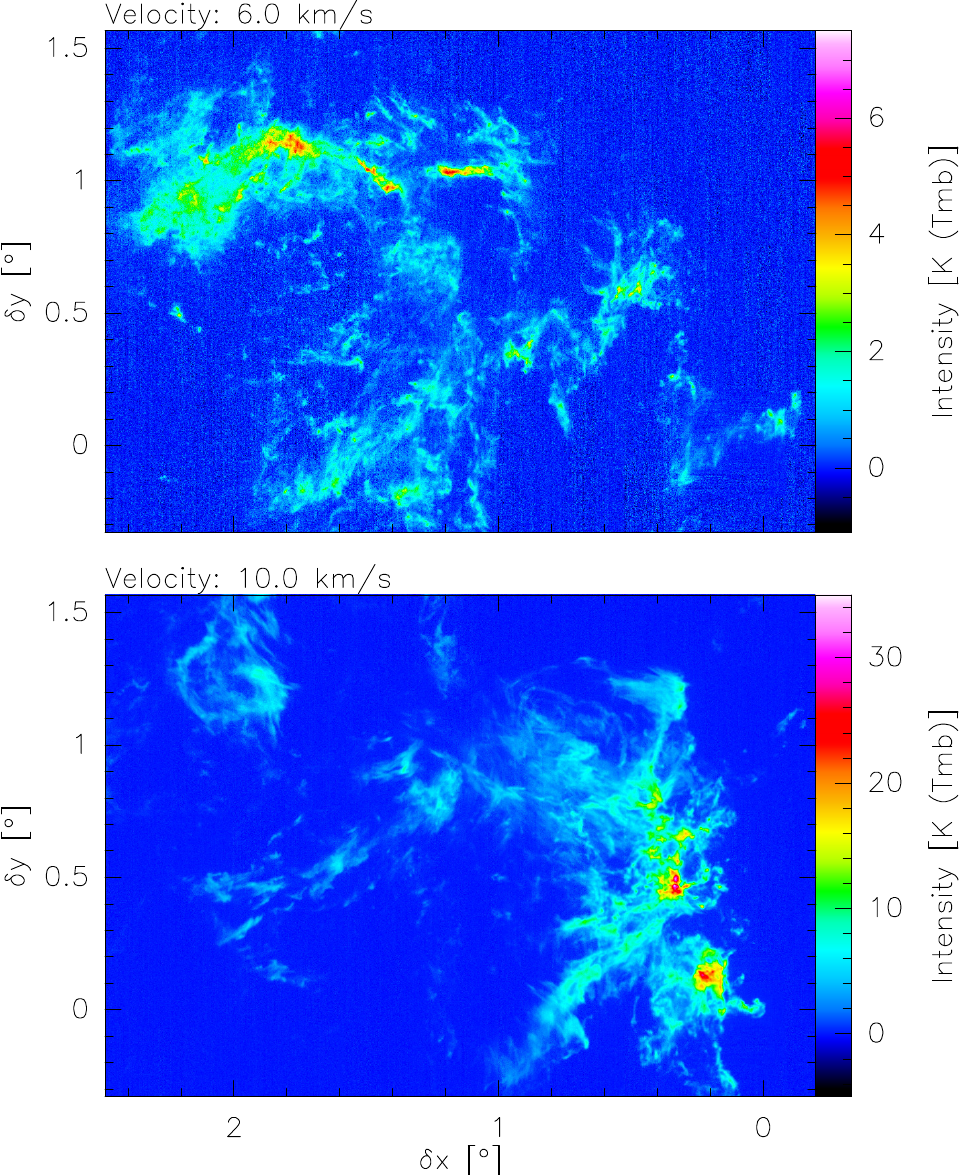}
    \caption{Velocity channels at 6 and 10 \kms{} of \thCO{} \Jone{}. The
      corresponding channels are displayed as vertical dashed red lines in
      Fig.~\ref{fig:data:spectra}.}
    \label{fig:data:selected:channels}
  \end{figure}
}

\newcommand{\FigDataCorrMI}{%
  \begin{figure}
    \centering %
    \includegraphics[width=\linewidth]{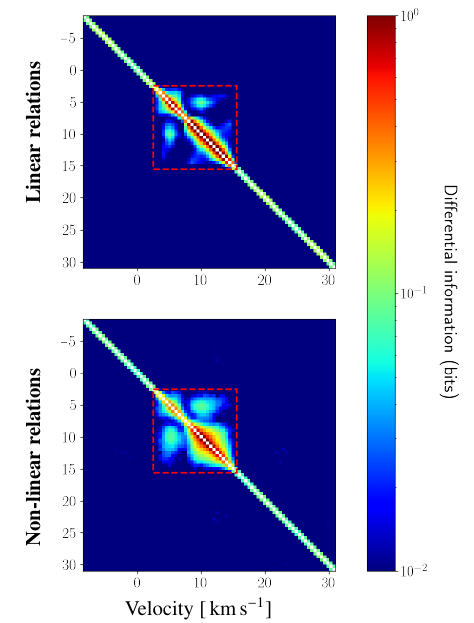}
    \caption{Amount of information shared between channels for the
      \thCO{}\Jone{} data cube. The \textbf{top} row shows information
      related only to linear relationship, while the \textbf{bottom} row
      shows information related to any type of relation (\ie, the mutual
      information).}
    \label{fig:data:corr:mi}
  \end{figure}
}

\newcommand{\FigDataNoiseSpatialPowerDensity}{%
  \begin{figure*}
    \centering %
    \includegraphics[width=\linewidth]{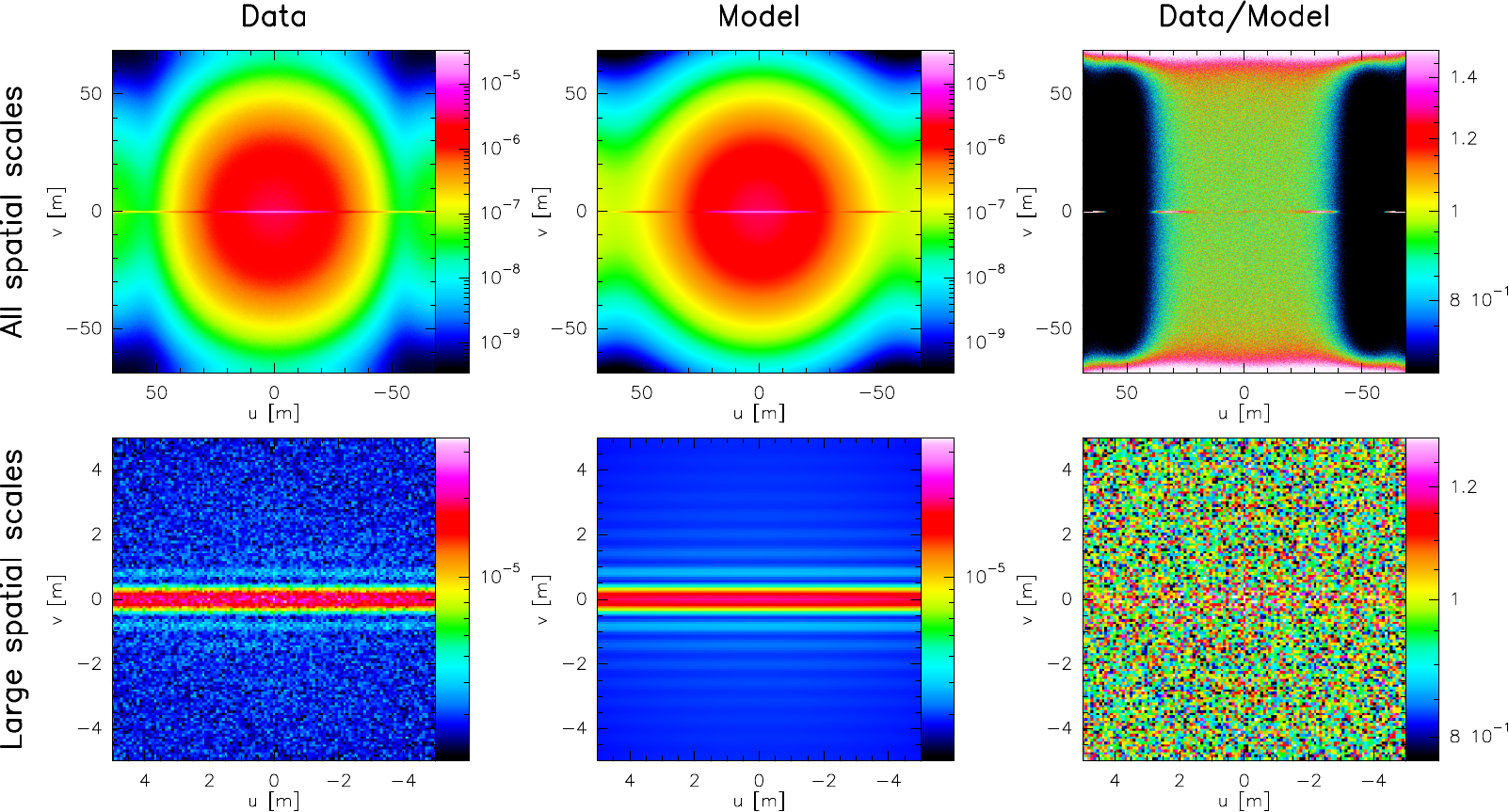}
    \caption{Comparison between the measured \textbf{(left)} and modeled
      \textbf{(middle)} noise spatial power density, and their ratios
      \textbf{(right)} in logarithmic scale. The top row shows the spatial
      power densities for all scales, while the bottom row zooms in on the
      large spatial scales.}
    \label{fig:data:noise:spatial:power:density}
  \end{figure*}
}

\newcommand{\FigDataNoiseSpectralPowerDensity}{%
  \begin{figure}
    \centering %
    \includegraphics[width=\linewidth]{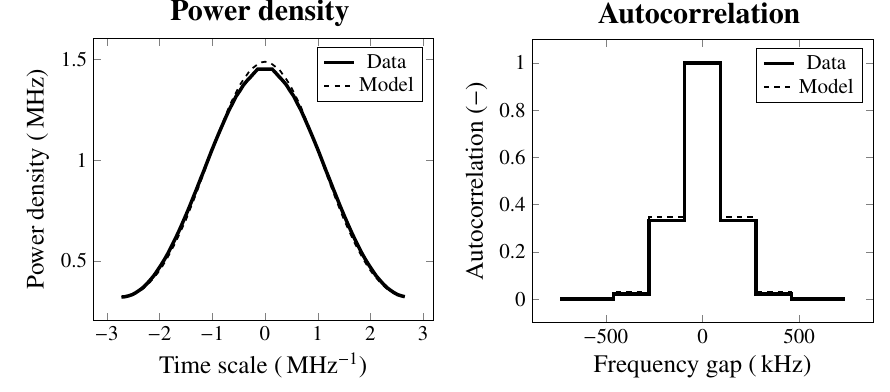} %
    \caption{Comparison between the measured \textbf{(plain line)} and
      modeled \textbf{(dashed line)} spectral power density \textbf{(left)}
      and autocorrelation function \textbf{(right)}.}
    \label{fig:spectral_power_density}
  \end{figure}
}

\newcommand{\FigNormes}{%
  \begin{figure}
    \centering
    \includegraphics[width=0.35\linewidth,angle=270]{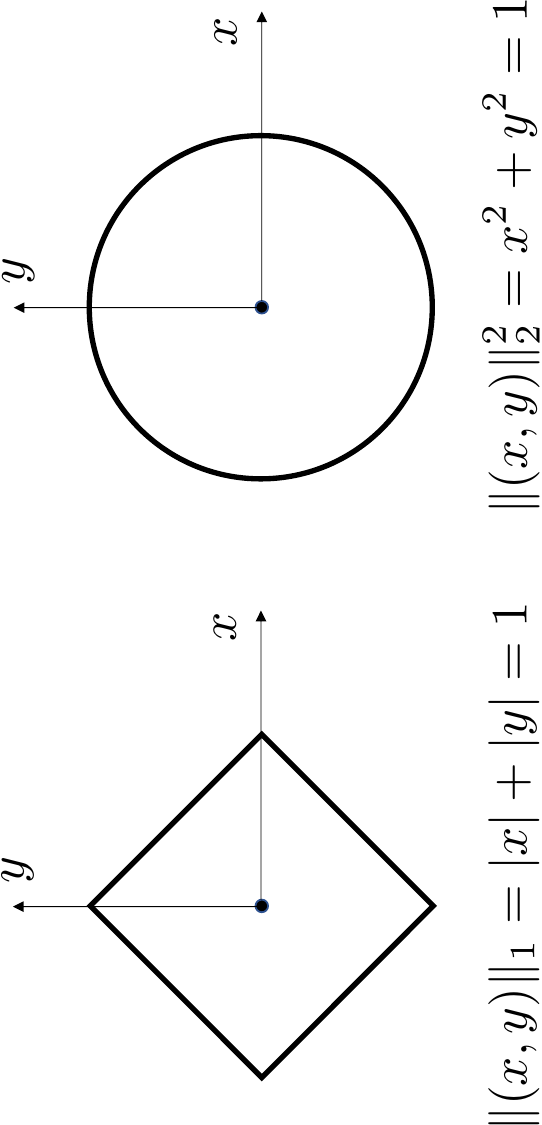}
    \caption{Illustration of the non-invariance to rotation of the $L_1$
      norm as opposed to the $L_2$ norm.}
    \label{fig:normes}
  \end{figure}
}

\newcommand{\FigStandardAutoencoder}{%
  \begin{figure}
    \centering %
    \includegraphics[width=\linewidth,trim={2cm 3cm 2cm 3cm}]{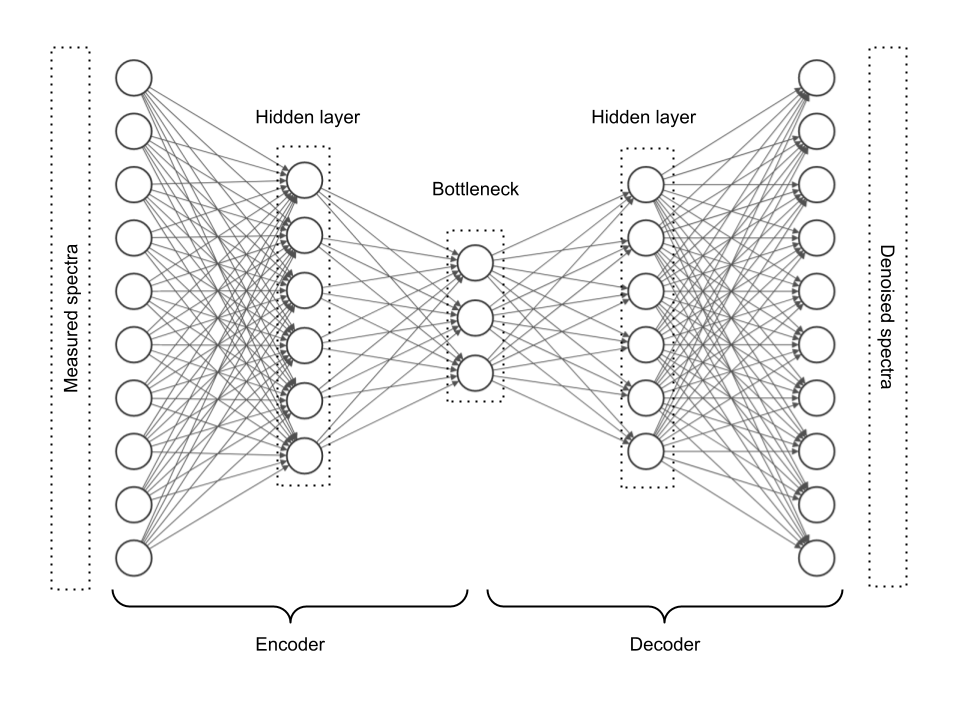}
    \caption{Example of an autoencoder neural network. Each column
      represents a neuron layer. Each arrow represents a connection between
      the neuron layers. The first and last layers are composed from the
      measured and denoised intensities of a spectrum at the different
      channels, respectively. The bottleneck contains the minimum number of
      neurons needed to compress the data without loss of signal
      information.  In this example, the signal intrinsic dimension (size
      of the bottleneck) is 3 while the data extrinsic one (size of the
      input and output spectra) is 10.}
    \label{fig:standard:autoencoder}
  \end{figure}
}

\newcommand{\FigDataIntrinsicDimensions}{%
  \begin{figure}
    \centering %
    \includegraphics[width=\linewidth]{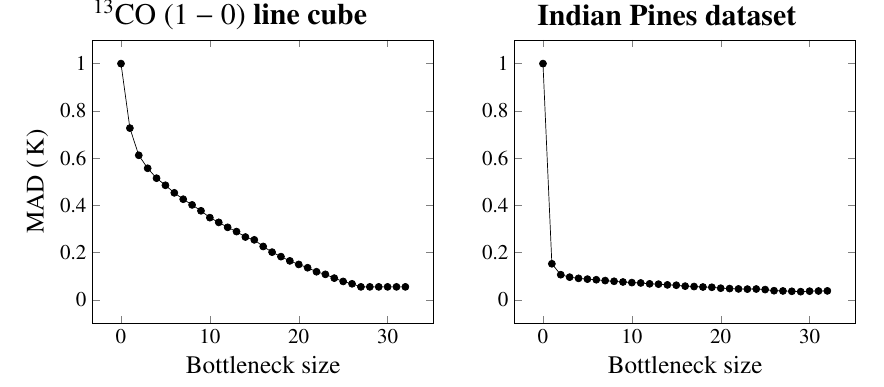}
    \caption{Distance (mean absolute deviation) between input and
      reconstructed data as a function of the bottleneck size for the
      \thCO{} \Jone{} data \textbf{(left)} and the Indian Pines data
      \textbf{(right)}.}
    \label{fig:data:intrinsic:dimensions}
  \end{figure}
}

\newcommand{\FigOptimizedAutoencoder}{%
  \begin{figure*}
    \begin{subfigure}{0.49\linewidth}
      \centering %
      \includegraphics[width=\linewidth,trim={0 4cm 0 2cm},clip]{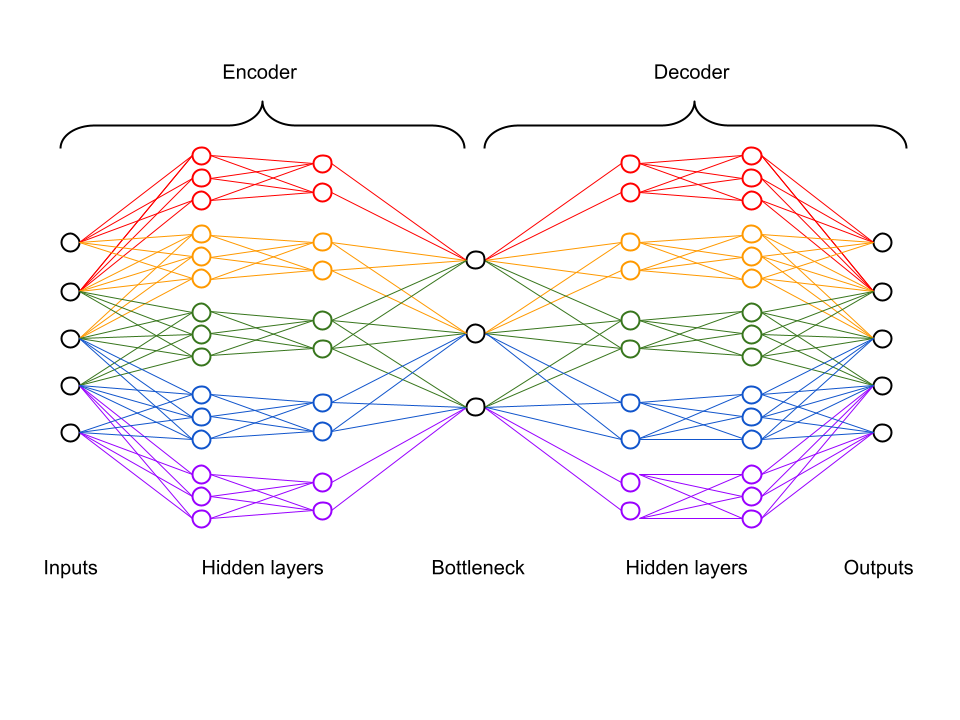}
      \caption{Optimized Autoencoder}
    \end{subfigure}
    \hfill%
    \begin{subfigure}{0.245\linewidth}
      \centering %
      \includegraphics[trim={9cm 0 9cm 1cm},clip,width=0.7\linewidth]{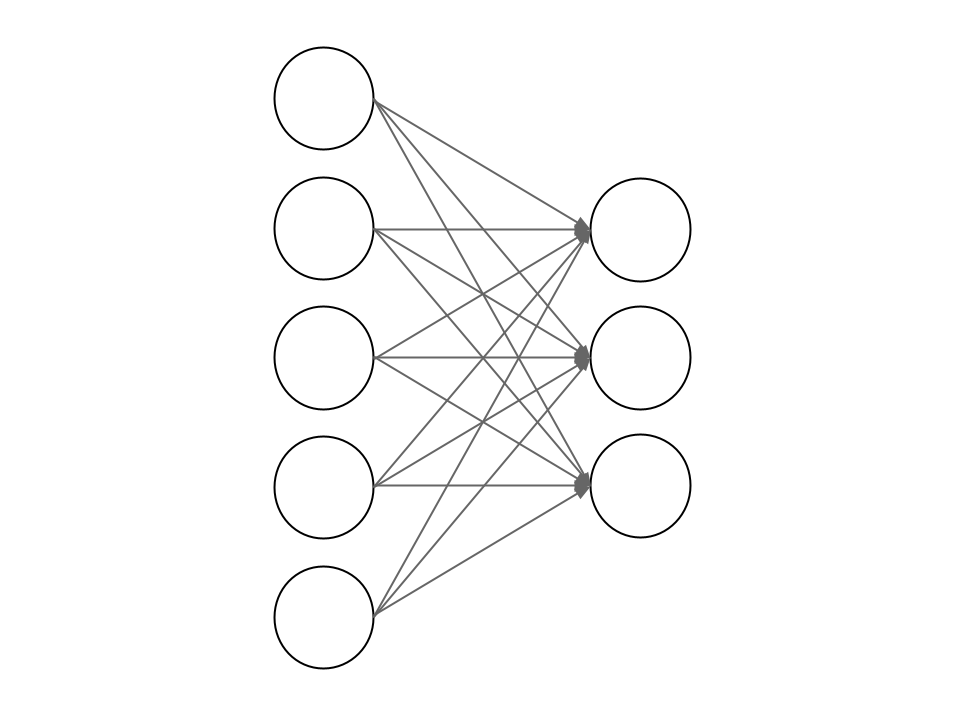}
      \caption{Fully connected layer}
    \end{subfigure}
    \hfill%
    \begin{subfigure}{0.245\linewidth}
      \centering %
      \includegraphics[trim={9cm 0 9cm 1cm},clip,width=0.7\linewidth]{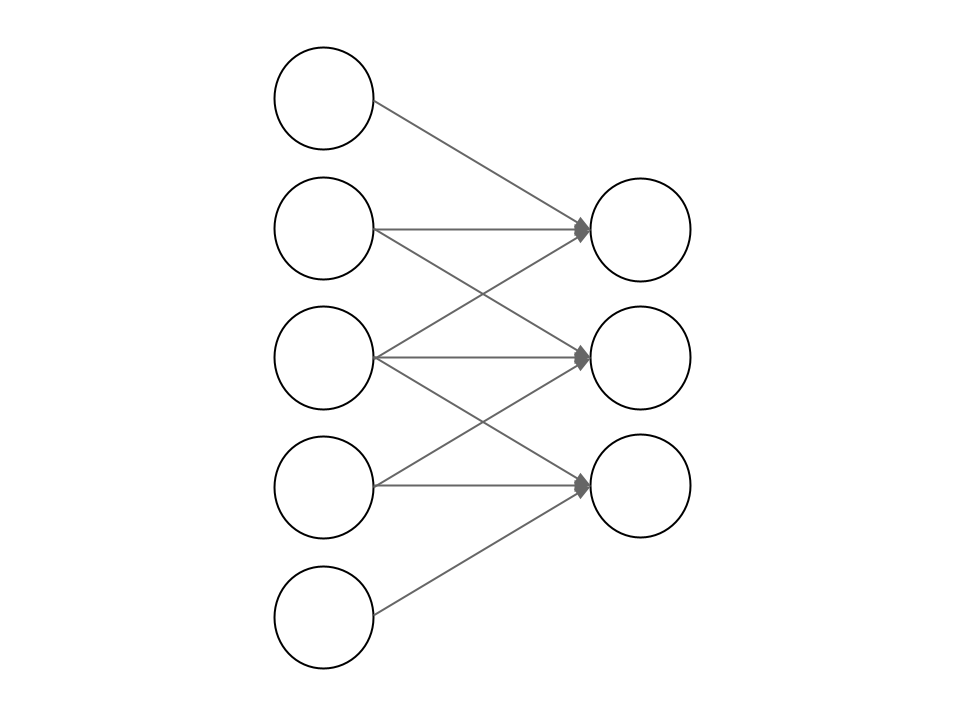}
      \caption{Locally connected layer}
    \end{subfigure}
    \caption{Optimized autoencoder architecture (\textbf{a}) where fully
      connected layers (\textbf{b}) are replaced by locally connected
      layers (\textbf{c}). The number of entries is 5 and the bottleneck is
      size 3. The hidden layers of the network can be described by
      describing the small encoders, here they are of dimension [3, 2] with
      input and output windows of the same size 3.}
    \label{fig:optimized:autoencoder}
  \end{figure*}
}

\newcommand{\FigDetectionSNR}{%
  \begin{figure*}
    \centering %
    \includegraphics[width=\linewidth]{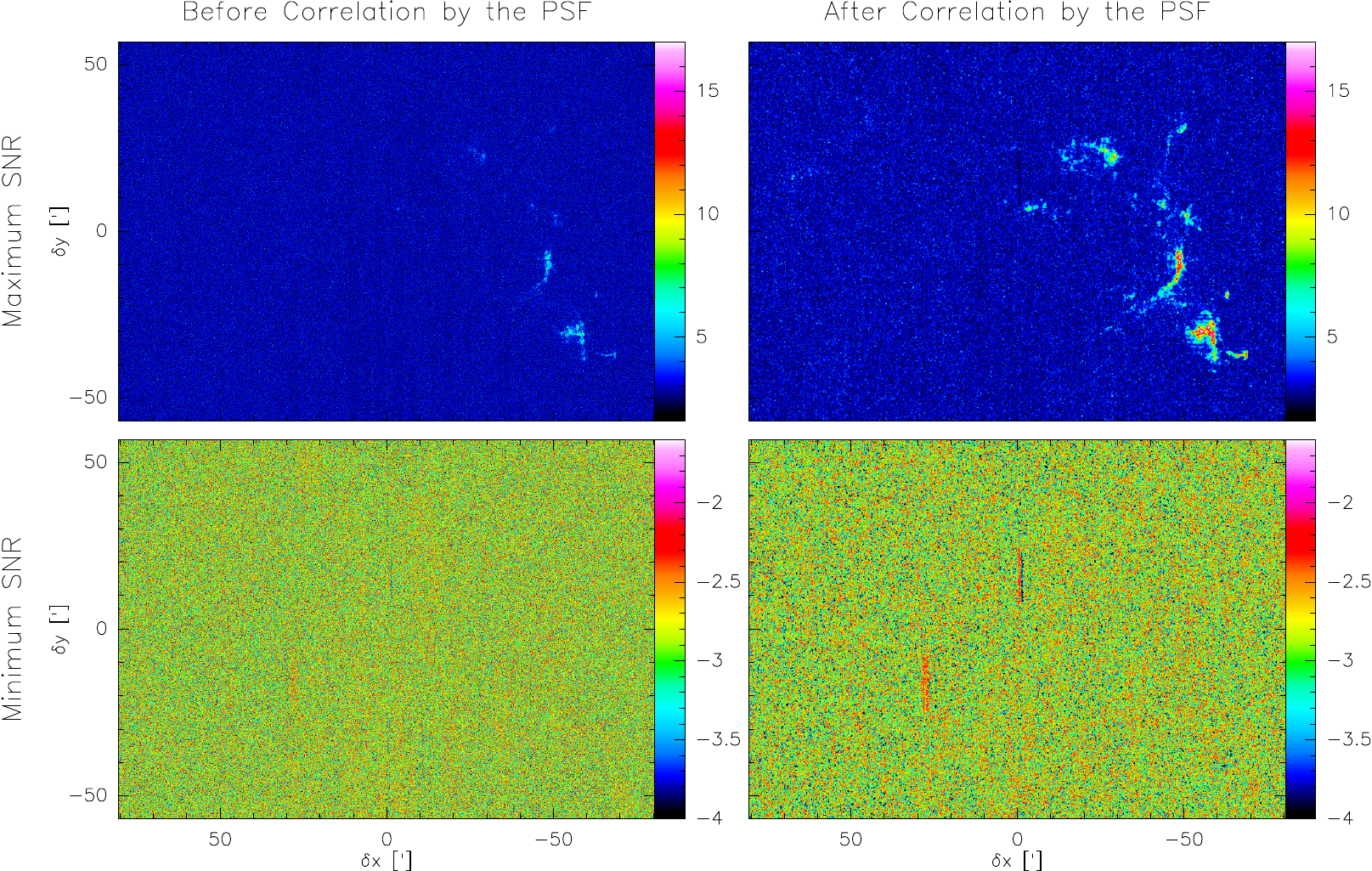}
    \caption{Maps of the maximum \textbf{(top)} and minimum
      \textbf{(bottom)} signal-to-noise ratio per spectrum before
      \textbf{(left)} and after \textbf{(right)} convolution of the \CseO{}
      \Jone{} line cube by the telescope point spread function.}
    \label{fig:detection:SNR}
  \end{figure*}
}

\newcommand{\FigDetectionSegments}{%
  \begin{figure*}
    \centering %
    \includegraphics[width=\linewidth]{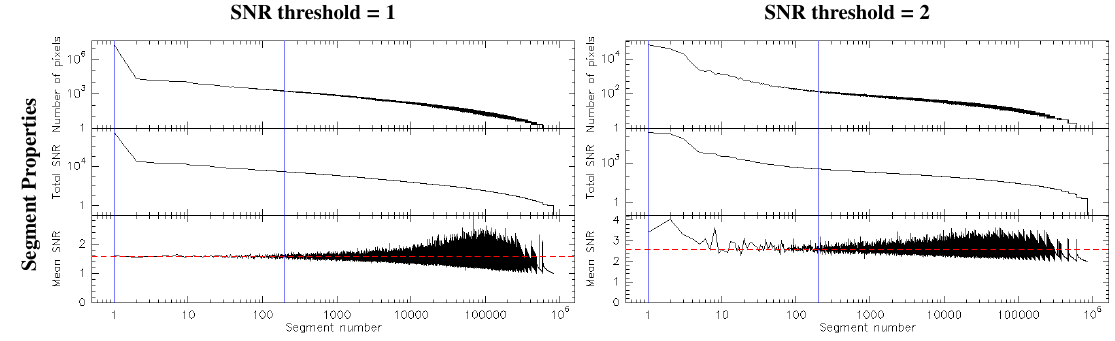}
    \caption{Properties of the segments obtained on the cube of
      signal-to-noise ratio (SNR) for the \CseO{} \Jone{} line. This cube
      was segmented into contiguous position-position-velocity regions
      above a minimum SNR value. The segments are ordered by decreasing
      value of the SNR summed over the segment (total SNR). The shown
      properties are, from top to bottom, the total number of pixels inside
      the segment, the total SNR, and the mean SNR of the segment. These
      properties are shown for two different SNR thresholds: 1 and 2. The
      blue plain vertical lines show the segments that are selected to
      compute the moment maps in Fig.~\ref{fig:detection:moments}. The red
      dashed horizontal lines show the typical mean SNR reached for the
      segment \# 200.}
    \label{fig:detection:segments}
  \end{figure*}
}

\newcommand{\FigDetectionMoments}{%
  \begin{figure*}
    \centering %
    \includegraphics[width=\linewidth]{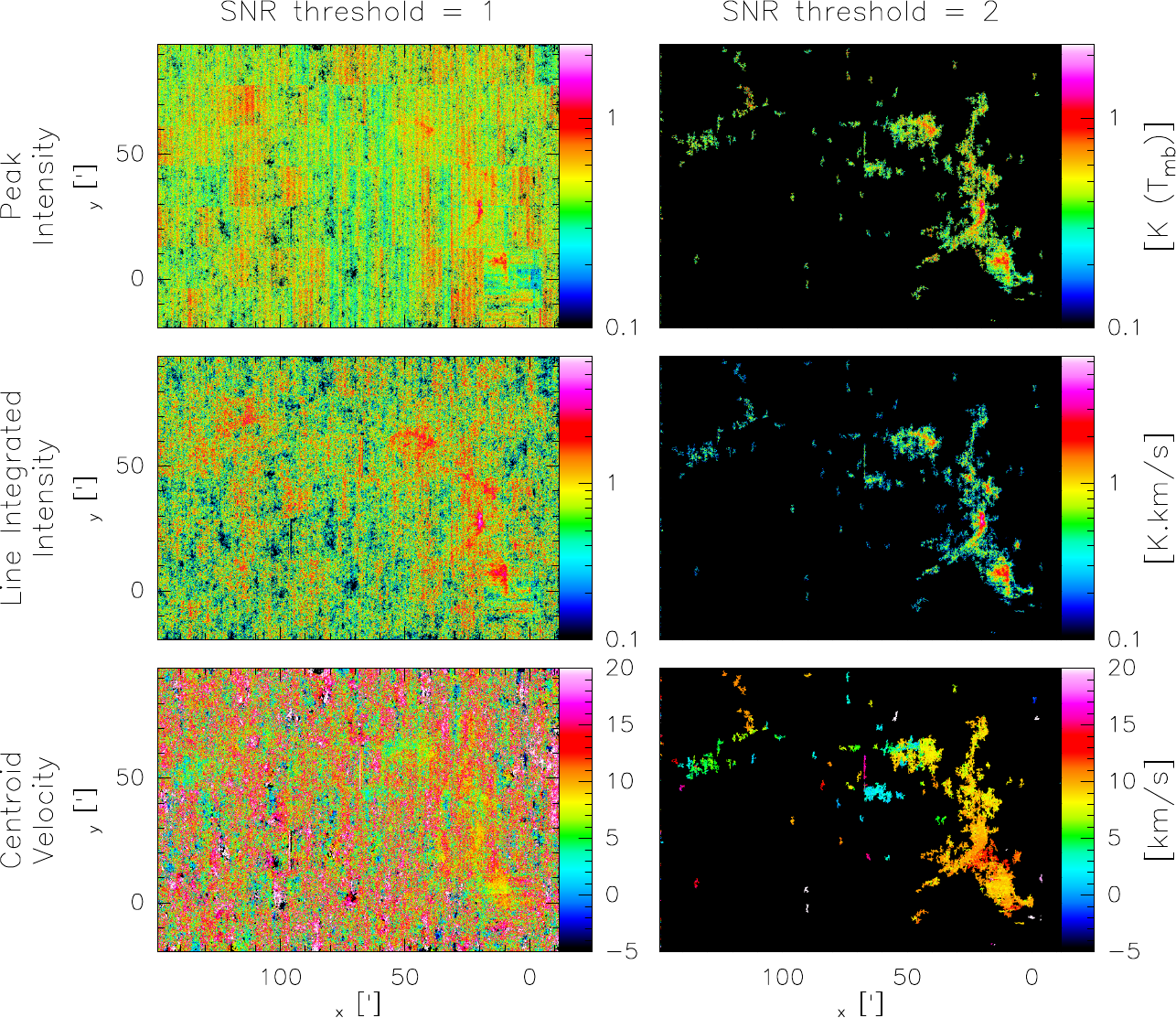}
    \caption{Maps of the moments of the spectrum for two different values
      (1 at left, and 2 at right) of the SNR threshold used to compute the
      position-position-velocity mask of significant emission. From top to
      bottom, the peak intensity (maximum of the spectrum), line integrated
      intensity (moment 0 of the spectrum), and centroid velocity (moment 1
      of the spectrum) are shown.}
    \label{fig:detection:moments}
  \end{figure*}
}

\newcommand{\FigChannelComparison}{%
  \begin{figure*}
    \centering
    \includegraphics[width=0.475\linewidth]{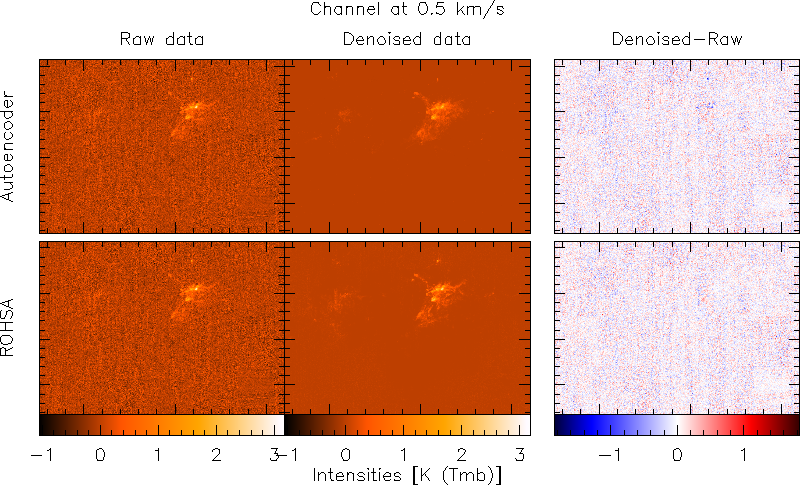}
    \hfill %
    \includegraphics[width=0.475\linewidth]{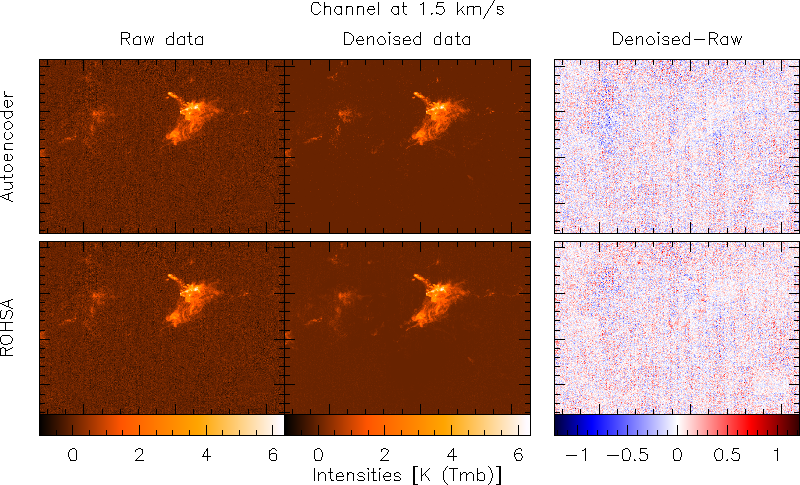}
    \\[\bigskipamount]
    \includegraphics[width=0.475\linewidth]{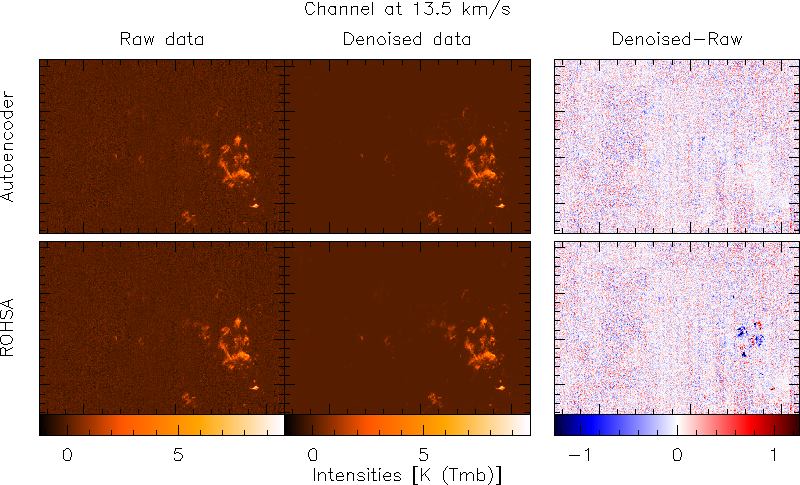}
    \hfill %
    \includegraphics[width=0.475\linewidth]{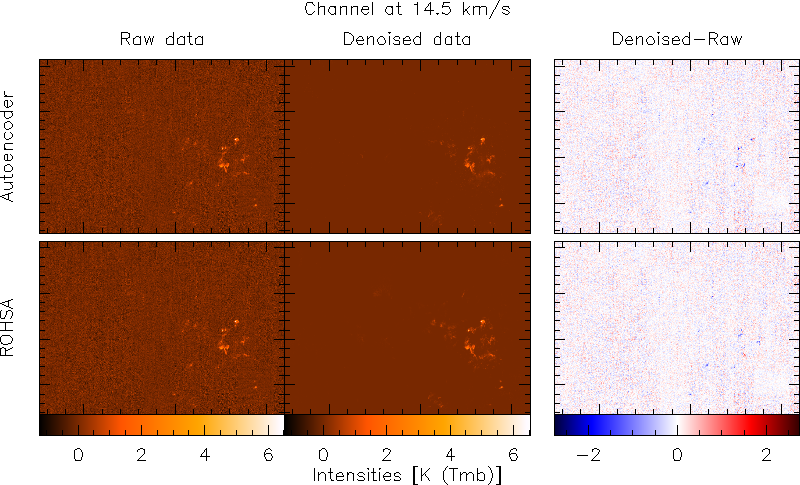}
    \caption{Comparison of the denoising performances of the taylored
      autoencoder and ROHSA for four different velocity channels belonging
      to the line wings. For each channel, the raw \textbf{(left)} and
      denoised \textbf{(middle)} images are shown with the same intensity
      scale and the residual \textbf{(right)} image is displayed with an
      optimized intensity scale. The top and bottom rows show the results
      for the autoencoder and ROHSA algorithms, respectively.}
    \label{fig:denoising:comparison:channels}
  \end{figure*}
}

\newcommand{\FigStatisticalComparisonDenoisingRMSE}{%
  \begin{figure*}
    \centering
    \includegraphics[width=0.9\linewidth]{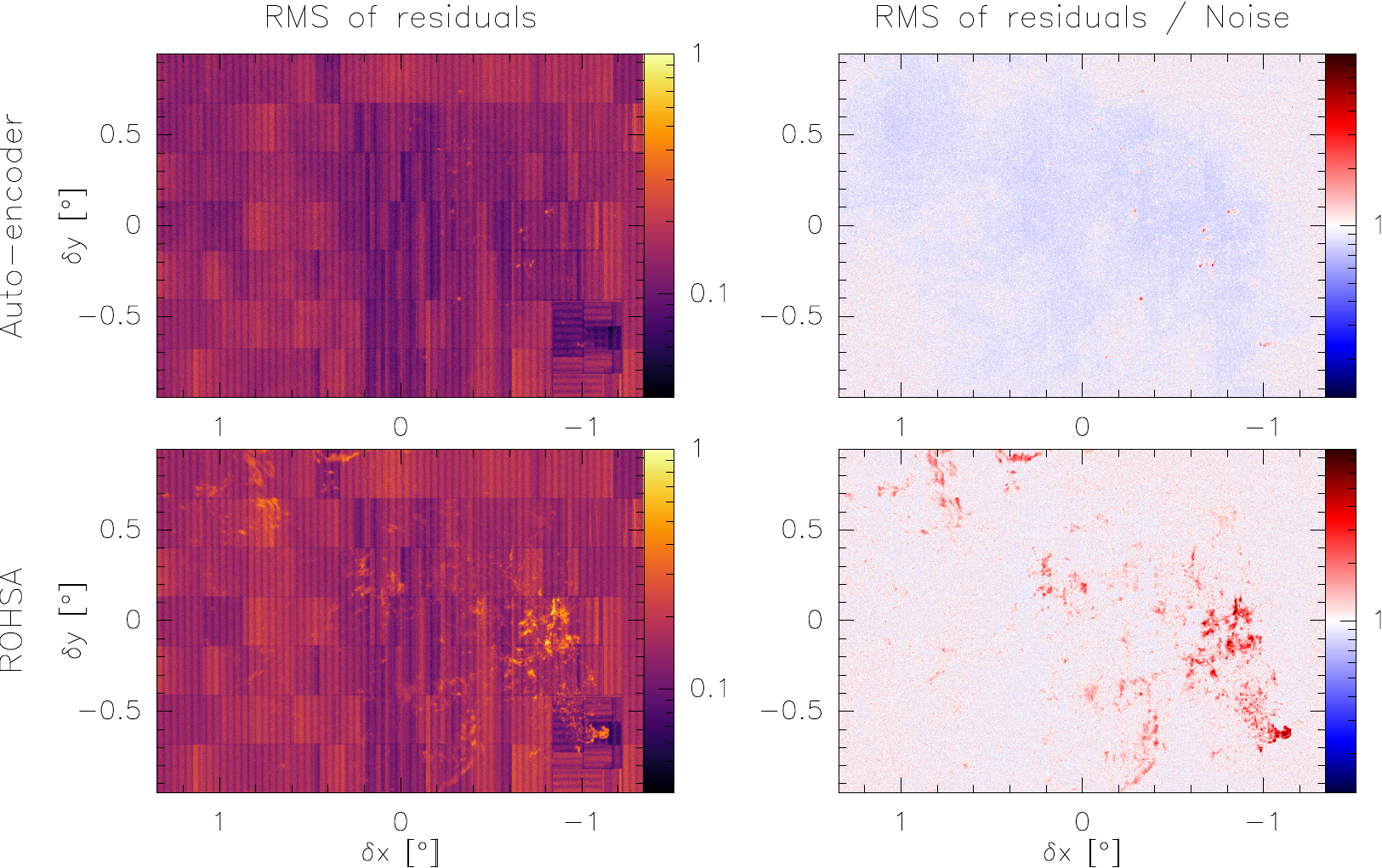}
    \caption{Comparison of the properties of the residuals after denoising
      by our autoencoder \textbf{(top)} and ROHSA \textbf{(bottom)}. The
      right column shows the map of the residual RMS, and the left column
      shows the map of the residual RMS normalized by the noise standard
      deviation.}
    \label{fig:denoising:comparison:rmse}
  \end{figure*}
}

\newcommand{\FigStatisticalComparisonDenoisingDenoisedVsRaw}{%
  \begin{figure}
    \centering
    \includegraphics[width=\linewidth]{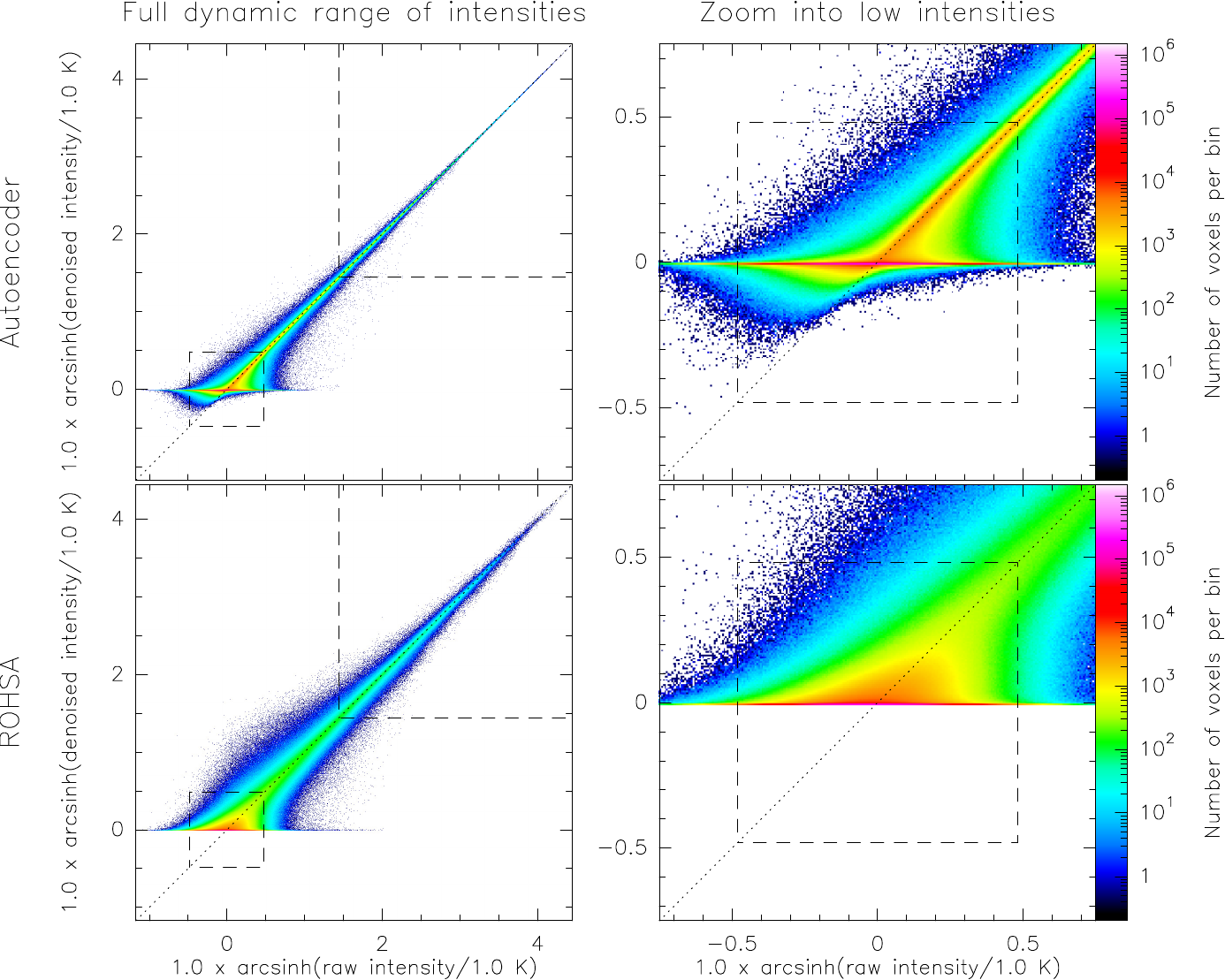}
    \caption{Comparison of the denoising performances of the taylored
      autoencoder \textbf{(top)} and ROHSA \textbf{(bottom)}. Each panel
      shows the joint histogram of the denoised intensities vs the data
      intensities. The left and right columns display the full dynamic
      range of intensities and a zoom into the low intensities. A arcsinh
      transform was applied in order to show the intensities below
      $5\sigma$ (lower dashed square) with a linear scale and above
      $20\sigma$ with a logarithm scale (upper dashed square). The dotted
      line highlights the identity function.}
    \label{fig:denoising:comparison:histo}
  \end{figure}
}

\newcommand{\FigStatisticalComparisonDenoisingMeanSpectra}{%
  \begin{figure}
    \centering
    \includegraphics[width=\linewidth]{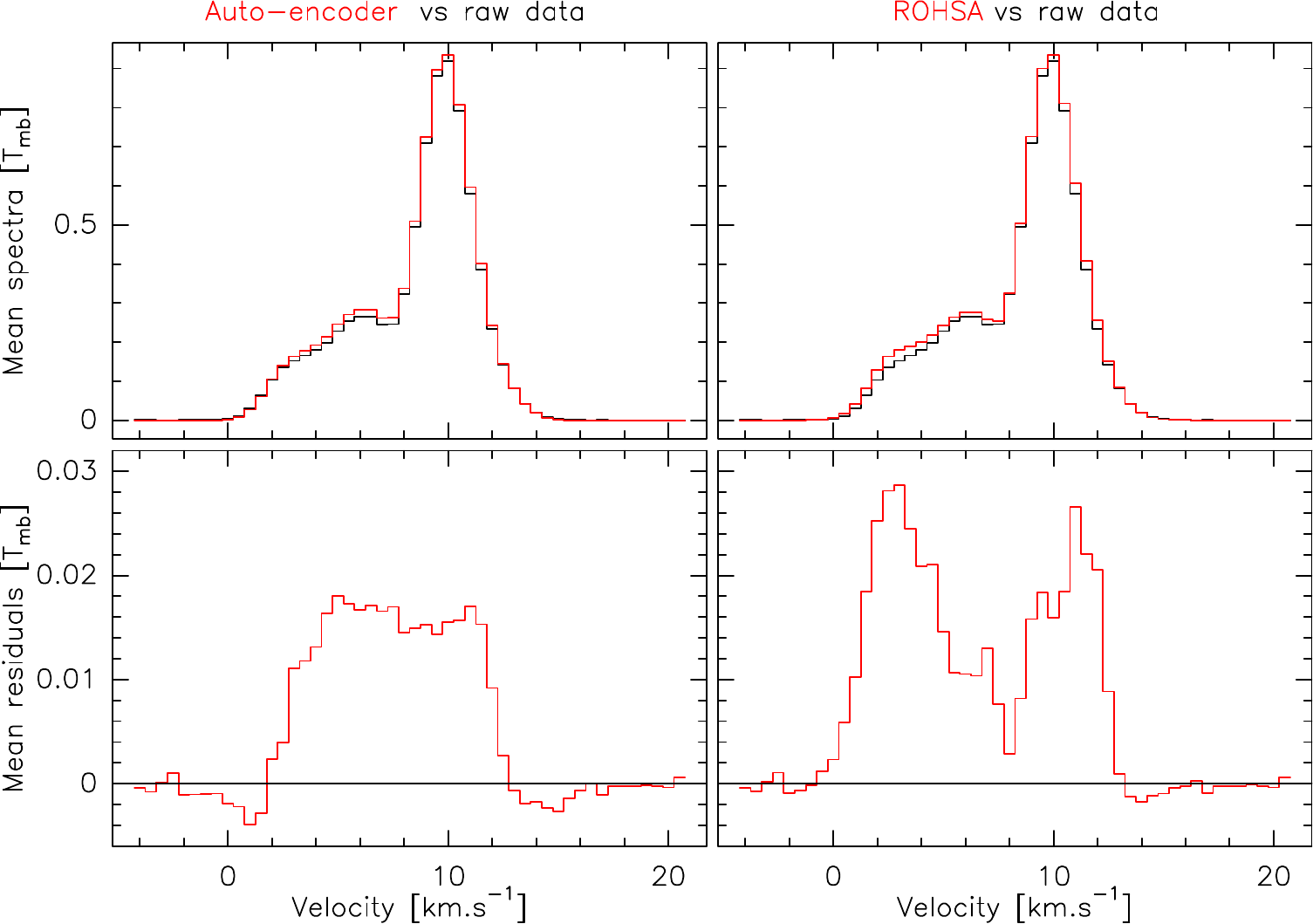}
    \caption{Comparison of the spectral profiles and residuals for the autoencoder \textbf{(left)} and ROHSA \textbf{(right)} algorithms. \textbf{Top:} Comparison of the input (in black) and output (in red) intensities. \textbf{Bottom:} Comparison of the residuals between the input and denoised data.}
    \label{fig:denoising:comparison:spectra}
  \end{figure}
}

\newcommand{\FigStatisticalComparisonDenoisingMomentsSignal}{%
  \begin{figure*}
    \centering
    \includegraphics[width=\linewidth]{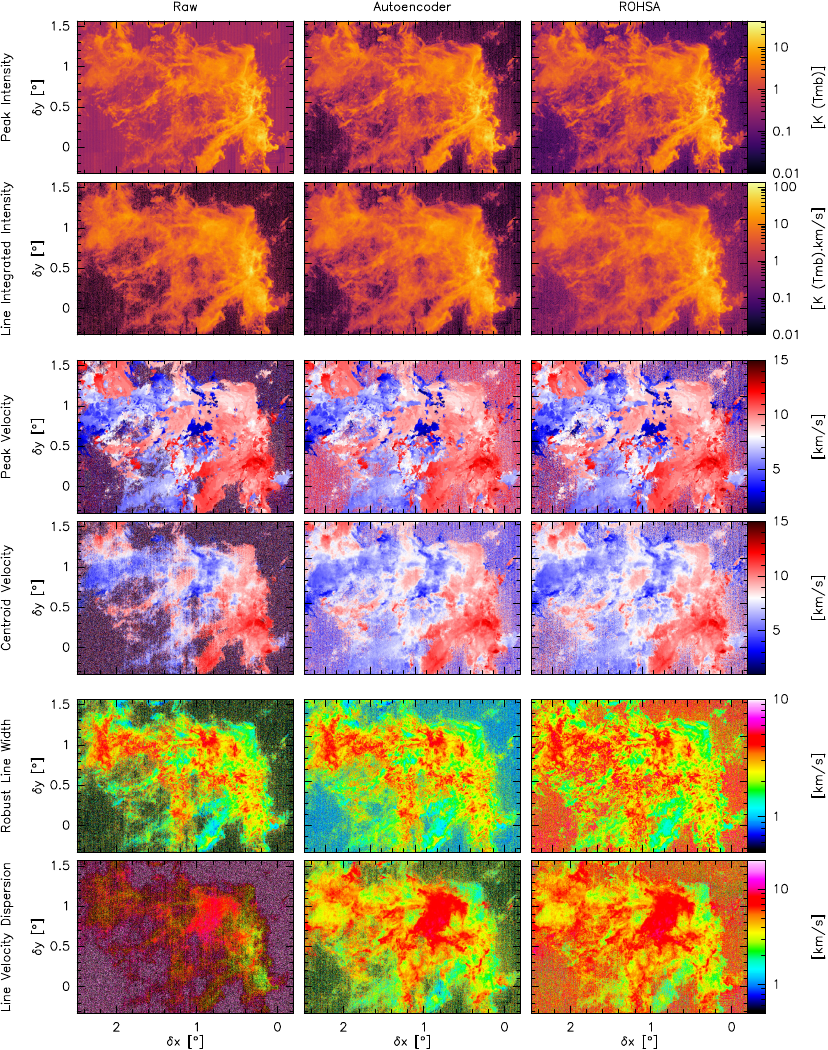}
    \caption{Maps of the properties of the \thCO{} \Jone{} line before
      \textbf{(left)} and after denoising with the autoencoder
      \textbf{(middle)} and ROHSA \textbf{(right)}. From top to bottom, the
      properties are the maximum of the line, the line integrated
      intensity, the velocity of the maximum, the centroid velocity, a
      robust estimation of the line width, and the velocity dispersion.}
    \label{fig:denoising:comparison:moments:signal}
  \end{figure*}
}

\newcommand{\FigStatisticalComparisonDenoisingMomentsDiff}{%
  \begin{figure*}
    \centering
    \includegraphics[width=\linewidth]{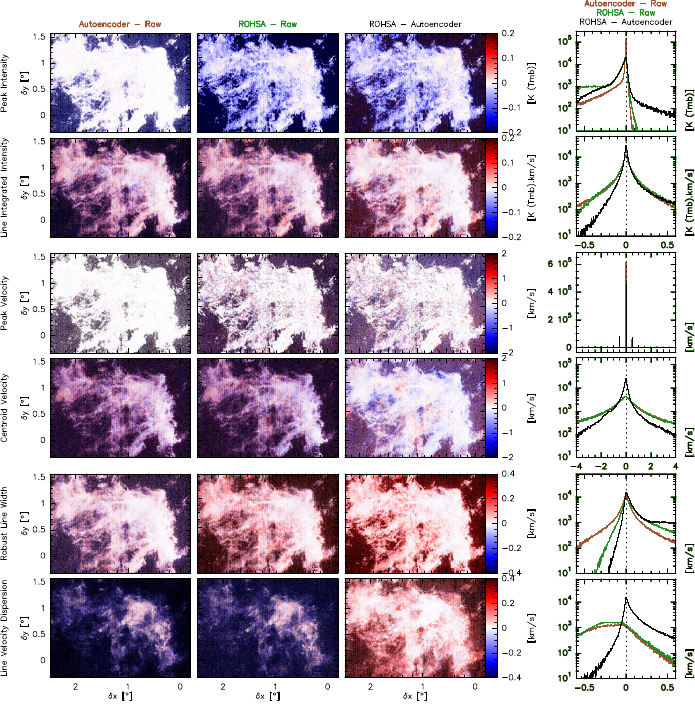}
    \caption{Maps and histograms of the residuals of the properties of the
      \thCO{} \Jone{} line. The properties are the same as for
      Fig.~\ref{fig:denoising:comparison:moments:signal}.  The first,
      second, and third columns show the maps of residuals between the
      autoencoder denoised and raw data, the ROHSA denoised and raw data,
      and the ROHSA and autoencoder denoised data, respectively. The color
      scales are saturated in order to emphasize the differences where some
      signal is detected. The fourth column shows the associated
      histograms. The brown and green lines show the residuals from the
      autoencoder and ROHSA denoising, respectively. The black lines show
      the difference between the autoencoder and ROHSA results.}
    \label{fig:denoising:comparison:moments:diff}
  \end{figure*}
}

\newcommand{\FigSchemaSpatialNoiseComputing}{%
  \begin{figure}[h]
    \centering %
    \includegraphics[width=\linewidth]{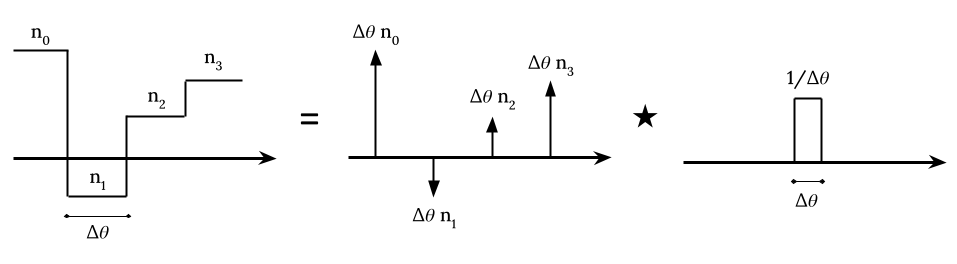}
    \caption{Illustration of the 1D calibration noise decomposition as the
      convolution between a random Dirac comb and a rectangular filter.}
    \label{fig:schema:spatial:noise}
  \end{figure}
}

\newcommand{\FigCalibrationUncertainty}{%
  \begin{figure}[h]
    \centering %
    \includegraphics[width=\linewidth]{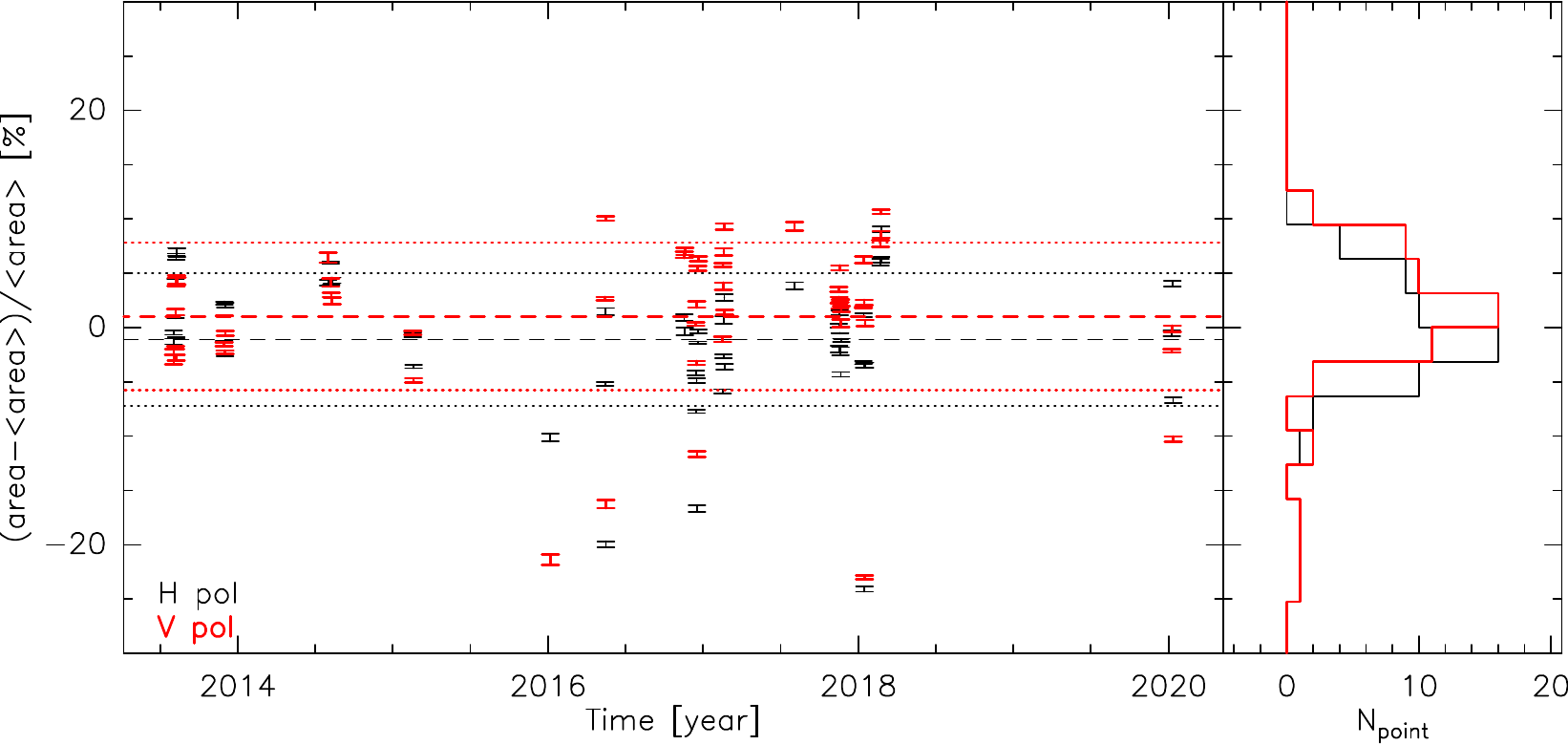}
    \caption{Relative variation of the fitted Gaussian area for the \thCO{}
      \Jone{} line towards the Horsehead core in percentage. The left panel
      shows the variations as a function of the time of the measurements,
      while the right panel shows the histogram of the variations. The
      black and red colors are used for the H and V polarizations of the
      EMIR receiver. The vertical error bars around each point show the
      uncertainties on the fitted area due to thermal noise. The horizontal
      dashed lines show the mean variation for each polarization. The
      horizontal dotted lines show the $\pm 1\sigma$ level for all the
      measurements.}
    \label{fig:data:calibration:uncertainty}
  \end{figure}
}

\newcommand{\FigPerceptron}{%
  \begin{figure}[h]
    \centering
    \includegraphics[width=0.9\linewidth]{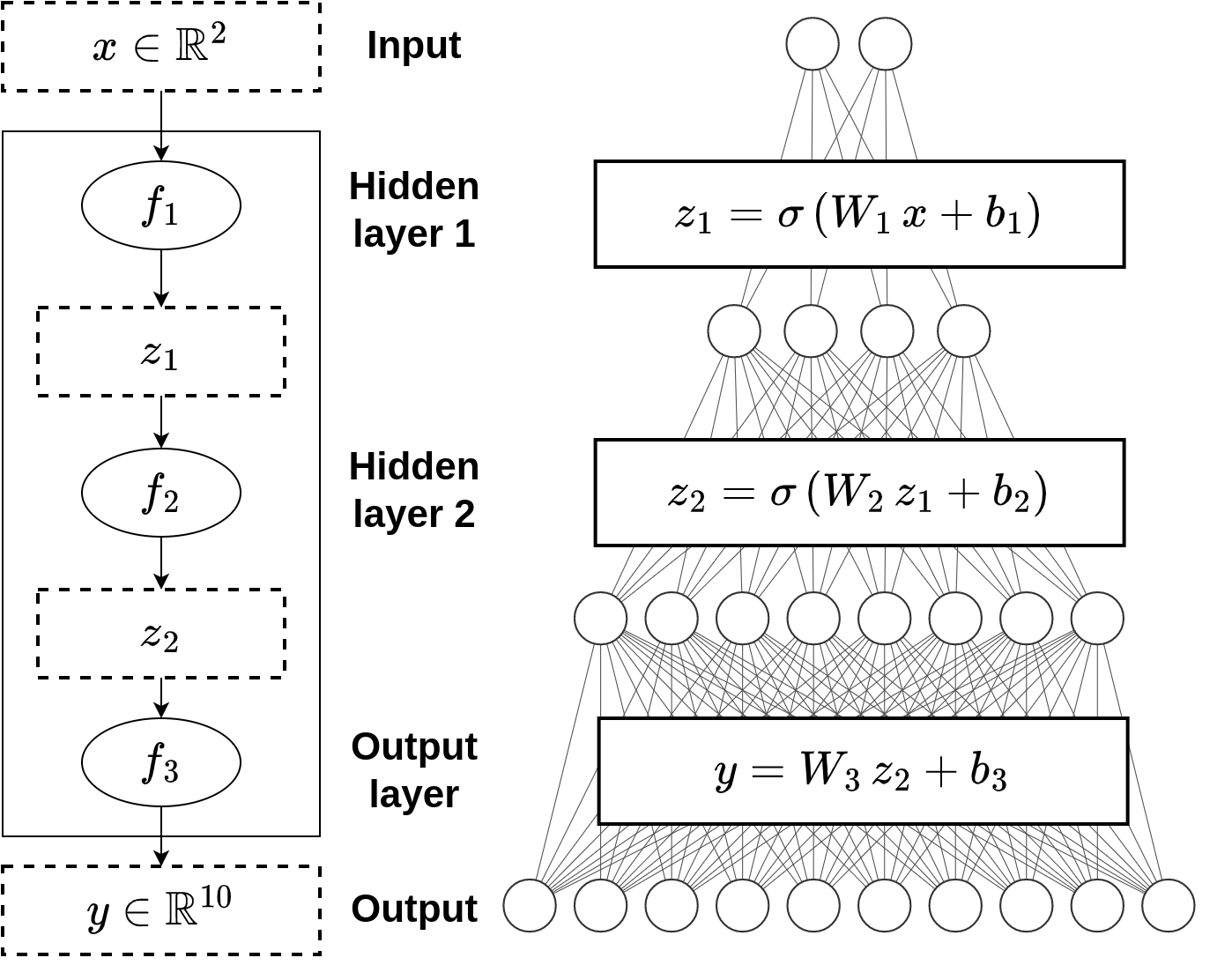}
    \caption{An example of multilayer perceptron with 2 inputs, 10 outputs
      and two hidden layers with 4 and 8 neurons, respectively.}
    \label{fig:perceptron}
  \end{figure}
}

\newcommand{\FigAllChannelsRadio}{%
  \begin{figure}[h]
    \centering %
    \includegraphics[width=\linewidth]{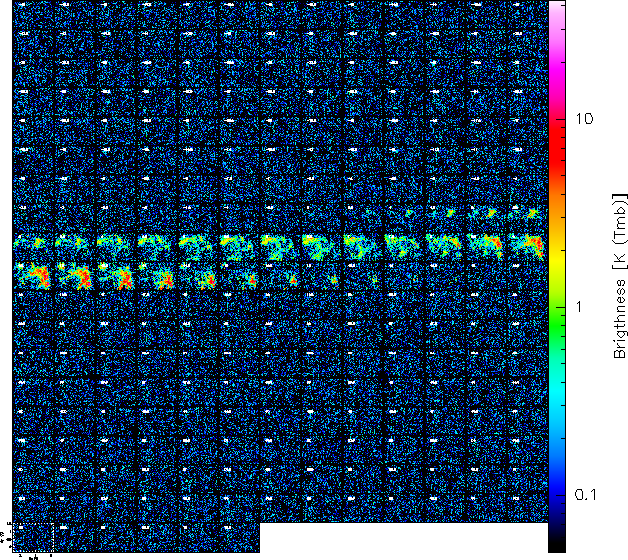}
    \caption{Visualization of the 240 spectral images of the \mbox{\thCO{}
      \Jone{}} cube sorted by increasing velocity.}
    \label{fig:all:channels:radio}
  \end{figure}
}

\begin{document}

\title{Deep learning denoising by dimension reduction: \\
  Application to the ORION-B line cubes}

\author{%
  Lucas Einig\inst{\ref{IRAM},\ref{GIPSA-Lab}}
  \and Jérôme Pety\inst{\ref{IRAM},\ref{LERMA/PARIS}} %
  \and Antoine Roueff\inst{\ref{Toulon}} %
  \and Paul Vandame\inst{\ref{GIPSA-Lab}} %
  \and Jocelyn Chanussot\inst{\ref{GIPSA-Lab}} %
  \and Maryvonne Gerin\inst{\ref{LERMA/PARIS}} %
  \and Jan H. Orkisz\inst{\ref{Chalmers}} %
  \and Pierre Palud\inst{\ref{CRISTAL},\ref{LERMA/MEUDON}}
  \and Miriam G. Santa-Maria\inst{\ref{CSIC}} %
  \and Victor de Souza Magalhaes\inst{\ref{IRAM}} %
  \and Ivana Be\v{s}li\'c\inst{\ref{LERMA/PARIS}}%
  \and S\'ebastien Bardeau\inst{\ref{IRAM}} %
  \and Emeric Bron\inst{\ref{LERMA/MEUDON}} %
  \and Pierre Chainais\inst{\ref{CRISTAL}} %
  \and Javier R. Goicoechea\inst{\ref{CSIC}} %
  \and Pierre Gratier \inst{\ref{LAB}} %
  \and Viviana V. Guzm\'an\inst{\ref{Catholica}} %
  \and Annie Hughes\inst{\ref{IRAP}} %
  \and Jouni Kainulainen\inst{\ref{Chalmers}} %
  \and David Languignon\inst{\ref{LERMA/MEUDON}} %
  \and Rosine Lallement\inst{\ref{GEPI}} %
  \and François Levrier\inst{\ref{LPENS}} %
  \and Dariusz C. Lis\inst{\ref{JPL}} %
  \and Harvey S. Liszt\inst{\ref{NRAO}} %
  \and Jacques Le Bourlot\inst{\ref{LERMA/MEUDON}} %
  \and Franck Le Petit\inst{\ref{LERMA/MEUDON}} %
  \and Karin \"Oberg\inst{\ref{CfA}} %
  \and Nicolas Peretto\inst{\ref{UC}} %
  \and Evelyne Roueff\inst{\ref{LERMA/MEUDON}} %
  \and Albrecht Sievers\inst{\ref{IRAM}} %
  \and Pierre-Antoine Thouvenin\inst{\ref{CRISTAL}} %
  \and Pascal Tremblin\inst{\ref{CEA}}
}

\institute{%
  IRAM, 300 rue de la Piscine, 38406 Saint Martin d'H\`eres,  France, \email{einig@iram.fr} \label{IRAM} %
  \and Univ. Grenoble Alpes, Inria, CNRS, Grenoble INP, GIPSA-Lab, Grenoble, 38000, France. \label{GIPSA-Lab} %
  \and LERMA, Observatoire de Paris, PSL Research University, CNRS, Sorbonne Universit\'es, 75014 Paris, France. \label{LERMA/PARIS} %
  \and Université de Toulon, Aix Marseille Univ, CNRS, IM2NP, Toulon, France. \label{Toulon} %
  \and Instituto de Física Fundamental (CSIC). Calle Serrano 121, 28006, Madrid, Spain. \label{CSIC} %
  \and Chalmers University of Technology, Department of Space, Earth and Environment, 412 93 Gothenburg, Sweden. \label{Chalmers} %
  \and Univ. Lille, CNRS, Centrale Lille, UMR 9189 - CRIStAL, 59651 Villeneuve d’Ascq, France. \label{CRISTAL} %
  \and LERMA, Observatoire de Paris, PSL Research University, CNRS, Sorbonne Universit\'es, 92190 Meudon, France. \label{LERMA/MEUDON} %
  \and Laboratoire d'Astrophysique de Bordeaux, Univ. Bordeaux, CNRS,  B18N, Allee Geoffroy Saint-Hilaire,33615 Pessac, France. \label{LAB} %
  \and Instituto de Astrofísica, Pontificia Universidad Católica de Chile, Av. Vicuña Mackenna 4860, 7820436 Macul, Santiago, Chile. \label{Catholica} %
  \and Institut de Recherche en Astrophysique et Planétologie (IRAP), Université Paul Sabatier, Toulouse cedex 4, France. \label{IRAP} %
  \and GEPI, Observatoire de Paris, PSL University, CNRS, 5 Place Jules Janssen, 92190 Meudon, France. \label{GEPI} %
  \and Laboratoire de Physique de l’Ecole normale supérieure, ENS, Université PSL, CNRS, Sorbonne Université, Université de Paris, Sorbonne Paris Cité, Paris, France. \label{LPENS} %
  \and Jet Propulsion Laboratory, California Institute of Technology, 4800 Oak Grove Drive, Pasadena, CA 91109, USA. \label{JPL} %
  \and National Radio Astronomy Observatory, 520 Edgemont Road, Charlottesville, VA, 22903, USA. \label{NRAO} %
  \and Harvard-Smithsonian Center for Astrophysics, 60 Garden Street, Cambridge, MA 02138, USA. \label{CfA} %
  \and School of Physics and Astronomy, Cardiff University, Queen's buildings, Cardiff CF24 3AA, UK. \label{UC} %
  \and AIM, CEA, CNRS, Université Paris-Saclay, Université Paris Diderot, Sorbonne Paris Cité, 91191 Gif-sur-Yvette, France. \label{CEA}%
} %


\abstract
{The availability of large bandwidth receivers for millimeter
  radio telescopes allows the acquisition of position-position-frequency
  data cubes over a wide field of view and a broad frequency
  coverage. These cubes contain much information on the physical, chemical,
  and kinematical properties of the emitting gas. However, their large size
  coupled with inhomogenous signal-to-noise ratio (SNR) are major
  challenges for consistent analysis and interpretation.}
{We search for a denoising method of the low SNR regions of the studied
  data cubes that would allow to recover the low SNR emission without
  distorting the signals with high SNR.}
{We perform an in-depth data analysis of the \thCO{} and \CseO{} \Jone{}
  data cubes obtained as part of the ORION-B large program performed at the
  IRAM 30m telescope. We analyse the statistical properties of the noise
  and the evolution of the correlation of the signal in a given frequency
  channel with that of the adjacent channels. This allows us to propose
  significant improvements of typical autoassociative neural
  networks, 
  often used to denoise hyperspectral Earth remote sensing data. Applying
  this method to the \thCO{} \Jone{} cube, we compare the denoised data
  with those derived with the multiple Gaussian fitting algorithm ROHSA,
  considered as the state of the art procedure for data line cubes.}
{The nature of astronomical spectral data cubes is distinct from that of the
  hyperspectral data usually studied in the Earth remote sensing literature
  because the observed intensities become statistically independent beyond
  a short channel separation. This lack of redundancy in data has led us to
  adapt the method, notably by taking into account the sparsity of the
  signal along the spectral axis. The application of the proposed algorithm
  leads to an increase of the SNR in voxels with weak signal, while
  preserving the spectral shape of the data in high SNR voxels.}
{The proposed algorithm that combines a detailed analysis of the noise
  statistics with an innovative autoencoder architecture is a promising
  path to denoise radio-astronomy line data cubes.  In the future,
  exploring whether a better use of the spatial correlations of the noise
  may further improve the denoising performances seems a promising
  avenue. In addition, dealing with the multiplicative noise associated
  with the calibration uncertainty at high SNR would also be beneficial for
  such large data cubes.}


\maketitle{} %


\section{Introduction}
\label{sec:introduction}

The current generation of millimeter radio-astronomy receivers is able to
produce large spectro-imaging data cubes (about $10^6$ pixels $\times 10^5$
frequencies or 0.4\,TB) at a sensitivity of $0.1\,$K (per pixel of
$\sim 9''\times 9''\times 0.5\kms$ in about 1000 hours of observing time
at, \eg, the IRAM 30m telescope~\citep{pety2017}. The next generation of
receivers will be between 25 and 50 times faster~\citep{pety2022}. Such
projects will thus move from the category of large programs, which are
difficult to carry out because they require more than 100 hours of
telescope time per semester, to typical programs that only ask for 20 to 40
hours per semester. Main challenges in interpreting these observations are
that 1) the noise level depends on the frequency, 2) the emission varies
from bright unresolved sources to faint extended ones, and 3) the intricate
gas kinematics of the emitting gas leads to complex emission line profiles
(non-Gaussian profiles, high velocity line wings, self-absorptions, etc.),
which vary from one pixel to other. Increasing the signal-to-noise ratio,
often referred to simply as denoising, is an important step to lead to new
discoveries by enlarging the space of achieved observing performances.

Denoising is an important topic in remote sensing, and many methods and
algorithms are found in the literature, for instance principal component
analysis~\citep[PCA, \eg,][]{wold1987principal},
kernel-PCA~\citep[\eg,][]{scholkopf1997kernel}, low rank tensor
decomposition~\citep[\eg,][]{harshman1970foundations} and total variation
methods~\citep[\eg,][]{vogel1996}. These methods try to compress and
uncompress the input data in a way that filters the noise but retain the
salient features of the signal. Among them, autoencoder neural networks are
interesting algorithms because they propose a generic nonlinear principal
component analysis, well adapted to hyperspectral data in Earth remote
sensing~\citep{licciardi2018}. We here explore the statistical nature of
signal and noise in millimeter radio-astronomy cubes in order to understand
the adaptations of typical autoencoders, which are required to efficiently
denoise these cubes.

This article is organized as follows.  Section~\ref{sec:denoising} presents
the general problem of denoising and the particular case of denoising by
dimension reduction. Section~\ref{sec:dataset} details the acquisition
processes that directly affect the properties of the
noise. Sections~\ref{sec:signal} and~\ref{sec:noise} characterize the
signal and noise properties for the studied line data cubes. The intrinsic
dimension of the signal is determined in
Sect.~\ref{sec:autoencoder:neural:networks}.
Section~\ref{sec:autoencoder:optimized} presents the modifications proposed
to typical autoencoder neural networks to better handle radio-astronomy
line cubes. The obtained denoising performances are then compared with the
state-of-the-art ROHSA algorithm in
Section~\ref{sec:denoising:performances}. Section~\ref{sec:conclusion}
summarizes the conclusions.


\section{Denoising by dimension reduction}
\label{sec:denoising}

\subsection{Definition of a denoising algorithm}
\label{sec:denoising:algo}

The observed data $d$ are noisy observations of the astronomical signal $s$
\begin{equation}
  d = f(s),
\end{equation}
where $f$ is a known function that describes the observing process with its
random component considered as noise. Denoising computes an estimate
$\hat{s}$ of the signal based on \textit{a priori} knowledge of the
deterministic and random part of the function $f$. This study will be
restricted to the case where the response $f$ of the telescope is linear
\begin{equation}
  d = c\cdot s+n,
\end{equation}
where $n$ is one realization of an additive random variable $N$, and $c$ is
one realization of a multiplicative random variable $C$. The variables $N$
and $C$ are centered on 0 and 1, respectively. In radio-astronomy, $N$
represents the thermal noise, and $C$ the calibration noise associated to
the uncertain determination of the calibration parameters (see
Sec.~\ref{sec:noise}). It is often assumed that the calibration uncertainty
is negligible. In this case, the performance of the denoising estimator can
be characterized by the improvement of the signal-to-noise ratio (SNR).

\subsection{Supervised vs self-supervised methods}
\label{sec:self:supervised}

In machine learning, denoising algorithms belong to two main categories.
\begin{description}
\item[\textbf{Supervised methods}] that use a set of known $(d,s)$ couples,
  called a training set, to train the algorithm to estimate $s$ from the
  measured values of $d$.  When available, ground truth data is the best
  choice to build the training set. In astrophysics, numerical simulations
  based on physical laws and laboratory experiments are used as
  surrogates. The simplifications required to be able to describe a
  complicated reality may bias the denoising.
\item[\textbf{Self-supervised methods}] consider that data are both the
  measurements (features) and ground truth (labels). Additional constraints
  on the denoising process are required to avoid delivering the data itself
  as the denoised estimate of the signal. A common assumption is that the
  signal $s$ is located in a lower dimension space than the observed data
  $d$. The idea is that the intrinsic dimension of the signal space is
  lower than its extrinsic dimension. For instance, let's assume that the
  data is composed of three features $(d_1,d_2,d_3)$ with 4 different
  samples for each of the feature, as in
  \begin{equation}
    [d_1,d_2,d_3] = \left[\begin{array}{rrr} 2&1&1\\1&-2&1\\5&6&4\\2&-8&4 \end{array}\right].
  \end{equation}
  The extrinsic dimension is 3, \ie, the number of features. But its
  intrinsic dimension is only 2. Indeed, the values of the features (\ie,
  the first, second and third columns of the above matrix) are
  deterministically linked to two independent variables $u$ and $v$ through
  \begin{equation}
    d_1 = u+v, \quad
    d_2 = uv, \quad \mbox{and} \quad
    d_3 = u^2,
  \end{equation}
  \begin{equation}
    \mbox{where} \quad
    [u,v] = \left[\begin{array}{rr} 1&1\\-1&2\\2&3\\-2&4 \end{array}\right].
  \end{equation}
  Any algorithm that will be able to deduce the above relations from the
  measured data will enable to compress it because only two numbers per
  sample are required to encode the three features. But it will also enable
  to denoise the data. Indeed, in presence of noise, knowing the
  relationship that exists between the features, will enable us to consider
  the measurement of the three features as three independent measurements
  of the same two underlying variables $u,v$, and thus to increase the
  signal-to-noise ratio of the estimated signal.
\end{description}

\subsection{Generic denoising by dimension reduction}
\label{sec:generic:denoising}

\subsubsection{Principle}
\label{sec:generic:denoising:principle}

Denoising by dimension reduction aims at mapping the data with an encoder
function ${\cal E} : \mathbb{R}^m \longrightarrow \mathbb{R}^l$ with $l<m$,
so that $\phi = {\cal E}(d)$ contains all the salient features $\phi$ of
the signal of interest $s$ and filters out the noise. The fact that $l<m$
implies that the encoder compresses the data.  Another function, named
decoder ${\cal D} : \mathbb{R}^l \longrightarrow \mathbb{R}^m$, estimates
the signal $s$ from its salient features without loss. The estimated signal
should preserve the relevant physical information from the astronomical
source, and it should have an increased SNR.  The spaces $\mathbb{R}^m$ and
$\mathbb{R}^l$ are thus called data and bottleneck (or latent) spaces,
respectively. The denoising will be all the better when $l \ll m$, and the
signal is extracted without distortion.

In astrophysics, denoising can be achieved with two different
approaches. First, astronomers may just wish to improve the signal-to-noise
ratio of the measurements to ease the extraction of the physical
information in a second step. The structure and unit of the estimated
signal stay unchanged. Second, astronomers may directly try to estimate the
physical parameters (\eg, the source geometry and kinematics, the volume
and column density, the kinetic temperature, the far-UV illumination, the
Mach number, the magnetic field, chemical abundances, etc), which best fit
the measured data. In this case, the significant physical and chemical
processes are selected, and their corresponding laws allow one to fit the
data. The salient features $\phi$ are the physical parameters of
interest. While this study will use the first approach, an interesting
challenge of denoising algorithms by dimension reduction is to enable
astrophysicists to relate the delivered salient features to the physical
quantities of interest. For instance, \citet{gratier2017} showed that the
first component of the principal component analysis of the integrated
intensities of a set of lines is related to the gas column density.

\subsubsection{In practice}
\label{sec:generic:denoising:practice}

Denoising by dimension reduction is thus based on a structure linking data $d$, estimated signal $\hat{s}$, and salient features $\phi$ as
\begin{equation}
  d(i_1,...,i_m) \xrightarrow{\cal E} \phi(j_1,...,j_l) \xrightarrow{\cal D} \hat{s}(i_1,...,i_m), %
  \quad \mbox{with} \quad %
  l < m.
\end{equation}
In principle, the level of distorsion should be measured as the distance
between $s$ and $\hat{s}$. However, it is impossible here because
astronomical observations of the interstellar medium do not provide ground
truth.  We will thus replace $\hat{s}$ by $s$ in the reminder of the paper
for the sake of simplicity.  In this representation, $(i_1,...,i_m)$ are
the spectral channels of the observed intensities, while $(j_1,...,j_l)$
are the indices of the salient features. The global denoising function
${\cal A}$, often called autoencoder, is defined as
\begin{equation}
  d \xrightarrow{\cal A={\cal D} \circ {\cal E}} s.
\end{equation}
It is just the composition of the ${\cal E}$ and ${\cal D}$ functions. The
function $\cal E$ and $\cal D$ are not exactly inverse of each
other. Indeed, in order to denoise, the function $\cal E$ must filter out
the noise. In other words, we expect that the function $\cal E$ will
transform a random variable $D$ of a large variance into a random variable
$\mathit{\Phi}$ of a low variance. There is no such requirement for the
function $\cal D$. For instance, denoising can sometimes be achieved
through the association of principal component analysis (PCA), which is a
linear inversible function, with a low dimensional projection.  After the
application of the PCA to the data, the components that better explain the
correlations of the original data are kept and the other ones are set to
zero, before inversing the PCA transformation. In this case, $\cal D$ is
the inverse of the PCA, while $\cal E$ is the PCA itself followed by a
nonlinear function that sets the noisiest (least informative from the
signal viewpoint) components to zero. In this case, the reduction of
dimensionality is obtained by enforcing a low dimensional bottleneck with
the direct transform before applying the inverse transform.

To achieve the denoising, it is necessary to estimate the best functions
$\cal E$ and $\cal D$ in terms of quality of reconstruction of the data for
a given dimensionality of the bottleneck space.
\begin{description}
\item[\textbf{Sampling the data}] Finding functions by numerical means
  first implies to correctly sample the manifold that links their input and
  output values. In other words, the algorithm must be trained with many
  (\eg, $K$) samples of the data $d$. This is subject to interpretation. In
  our case, the data is one position-position-channel cube
  $d(i_x,i_y.i_c)$, where $i_x,$ $i_y,$ and $i_c$ are the position of a
  pixel along the position and channel axes. This data cube can be seen as
  a set of images $d^\emr{ima}_{i_c}(i_x,i_y)$, or a set of spectra
  $d^\emr{spe}_{i_x,i_y}(i_c)$. The molecular line profiles are broadened
  by the gas motions along the line of sight. Optically thin lines deliver
  an approximation of the probability distribution function (PDF) of the
  velocity component parallel to the line of sight. As the interstellar
  medium is highly turbulent, the different spectra of one cube can be seen
  as the PDFs of many realizations of the underlying turbulent velocity
  field. This is the viewpoint used in this article.
\item[\textbf{Measuring the distance between $s$ and $d$ over all the
    samples}] Our goal is to find a single pair of functions
  $({\cal E},{\cal D})$ that correctly autoencodes all the samples of the
  data (all the spectra in our case). The distance between $s$ and $d$ is
  quantified with the mean squared error (MSE) between $d$ and $s$
  over all the samples
  \begin{equation}
    \emr{MSE}(s,d) = \frac{1}{K} \, \sum_{k=1}^K \paren{s_k-d_k}^2.
  \end{equation}
  The denoising problem can then be recast as an optimization problem whose
  goal is to find the function $\cal{A}$ that will minimize the distance
  between $s = {\cal A}(d)$ and $d$, \ie,
  \begin{equation}
    \label{eq:optimization:1}
    \hat{\cal A} = \arg\min_{\cal A} {\cal L}({\cal A},d_k),
  \end{equation}
  \begin{equation}
    \label{eq:loss:1}
    \mbox{with} \quad %
    {\cal L}({\cal A},d) = \frac{1}{K} \, \sum_{k=1}^K \bracket{{\cal A}(d_k)-d_k}^2.
  \end{equation}
  $\cal L$ is often called the loss function.
\end{description}
We now need to define the family of functions from which $\cal A$ will be
selected. Several ways can be used to reach this goal.
\begin{description}
\item[\textbf{Using generic function approximators}] as, \eg, neural
  networks.This will be our choice in this paper (see
  Sect.~\ref{sec:autoencoder:neural:networks}).
\item[\textbf{Using specific classes of function}] For instance,
  \citet{marchal2019} propose to fit the spectra as a finite set of
  Gaussian functions whose parameters (amplitude, position, full width at
  half maximum) can be spatially regularized. This methods is named ROHSA
  that stands for Regularized Optimization for Hyper-Spectral Analysis. In
  this case, $\cal D$ is a sum of Gaussians, $\cal E$ is the fitting
  algorithm, and the loss function is regularized as
  \begin{equation}
    {\cal L}({\cal A},d) = \emr{MSE}({\cal A}(d), d) + \frac{1}{K} \, \sum_{k=1}^K {\cal R}(k),%
    \quad \mbox{with} \quad %
  \end{equation}
  \begin{equation}
    {\cal R}(k) =
    \sum_{g=1,G}\,\cbrace{%
      \lambda_a      \norm{2}{{\cal K}\ast a_g}^2 + %
      \lambda_\mu    \norm{2}{{\cal K}\ast \mu_g}^2 + %
      \lambda_\sigma \norm{2}{{\cal K}\ast \sigma_g}^2}, %
  \end{equation}
  where $\cal K$ is a 2D convolution kernel that computes the second order
  differences, and $\lambda_a$, $\lambda_\mu$, and $\lambda_\sigma$ are the
  Lagrangian multipliers associated with convolved images of the amplitudes
  $a_g$, positions $\mu_g$, and standard deviations $\sigma_g$ of the $G$
  Gaussian functions. The value of the these multipliers needs to be fixed.
\end{description}


\section{Acquisition of radio-astronomy spectral line cubes by a
  ground-based single-dish telescope}
\label{sec:dataset}

\TabDataLines{} %

A detailed analysis of the radio-astronomical data is of critical
importance to understand the specificities of the considered data and thus
propose adequate optimizations for the denoising autoencoder.  To do this,
we first describe the acquisition of the data in detail to emphasize all
the phenomena that will impact the properties of the recorded signal and
noise.

\subsection{The ORION-B IRAM 30m Large Program}

The ORION-B project (Outstanding Radio-Imaging of \mbox{OrioN-B}, co-PIs:
J. Pety and M. Gerin) is a large program of the IRAM 30\,meter telescope
that aims to improve our understanding of physical and chemical processes
of the interstellar medium by mapping about half of the Orion B molecular
cloud over $\sim 85\%$ of the 3\,mm atmospheric window. The ORION-B field
of view covers five square degrees at a typical angular resolution of
$27''$ (or $50\mpc$ at a distance of 400\pc), or about $8\e{4}$ independent
lines of sight.

It uses the EMIR heterodyne receivers~\citep{carter2012} coupled with the
Fourier Transform Spectrometers~\citep{klein2006,klein2012} that
instantaneously deliver two spectra per polarization of 7.8\,GHz-bandwidth
sampled every 195\,kHz. These two spectra, named lower and upper
side-bands, are separated by 7.9\,GHz. The local oscillator of the
heterodyne receiver can be tuned at 3\,mm from $82.0$ to $107$\,GHz. This
enables a frequency coverage ranging from $70.7$ to $118.3$\,GHz in a few
successive observations. Moreover, the horizontal and vertical polarizations
are recorded and averaged. This delivers the total intensity of the source
(independent of the polarization state). It also allows us to gain a factor
of two on the acquisition time compared to recording a single polarization
state and assuming that the signal is unpolarized.

The ORION-B large program delivers a total bandwidth of about 40\,GHz at a
channel spacing of $\delta f = 195\kHz$, \ie, about 200\,000 channels. The
spectral resolving power (defined as $f/\delta f$, where $f$ is the
observing frequency) increases from $3.6\e{5}$ to $6.0\e{5}$ with
increasing frequency in the 3\,mm wavelength range. This huge resolving
power allows radio-astronomers to resolve the profiles from emission lines
of chemical tracers of the molecular gas, for instance, the \mbox{\J10}
lines of the isotopologues of carbon monoxide: \twCO{}, \thCO{}, \CeiO{},
and \CseO{}.

\subsection{Scanning strategy}

The heterodyne receivers currently available at the IRAM 30 meter telescope
can only record the emission towards a single direction of the sky at any
time. They are thus called single-beam receivers. To make an image with
such a detector, we need to scan the sky at a constant angular velocity
along lines of constant right ascension or declination. The signal is
continuously recorded and dumped at regular time intervals. This observing
mode is called On-The-Fly observations.

The data consist of a set of spectra that cover the target field of view in
a set of parallel lines. The angular distance $(\Delta\theta)$ between the
lines is set to satisfy the Nyquist sampling criterion
\begin{equation}
  \Delta \theta = \frac{\lambda}{2\,D},
\end{equation}
where $\lambda$ is the smallest observed wavelength, and $D$ is the
single-dish telescope diameter (30\,m here).

The resulting telescope response is slightly elongated along the scanning
direction because it is convolved along this direction with a boxcar filter
whose size corresponds to the angular size scanned during the integration
time~\citep{mangum2007}. To minimize this effect, it is desirable that the
telescope has moved only by a small fraction of its natural response during
one integration. We choose to dump the data 5 times over the angular scale
corresponding to the telescope natural beamwidth
\begin{equation}
  \theta = 1.2\,\frac{\lambda}{D}.
\end{equation}
We use the minimum sampling time that the computer system is able to
sustain during the typical duration of an observing session, \eg, 8
hours. With a dump time of 0.25 seconds, a scanning speed of $17''/$s
ensures a sampling of 5 dumps per beam along the scanning direction at the
$21.2''$ resolution reached at the highest observed frequency for the used
tuning, \ie, 116\GHz. The spatial sampling rates along and across the
scanning direction are adapted to the highest frequencies of each
individual tunings.

Only one scanning direction per tuning was observed in order to maximize
the observed field of view in the allocated telescope time. The usual
redundancy between horizontal and vertical scanning coverages could thus
not be exploited to improve the denoising algorithm.

\subsection{Calibration}
\label{sec:calibration}

Appendix~\ref{app:calibration} describes the methods used to calibrate the
data. Under perfect conditions, the calibrated spectrum, $S_\emr{cal}$, can
be written as
\begin{equation}
  S_\emr{cal}(f,\theta_l,\theta_m) = T_\emr{sys}\paren{f,\theta_l,\theta_m,\theta_{l0},\theta_{m0}}\,\cbrace{\frac{\emr{ON}(f,\theta_l,\theta_m)}{\emr{REF}(f,\theta_{l0},\theta_{m0})}-1},
\end{equation}
where $T_\emr{sys}(f)$ is the system temperature during the observation,
$\emr{ON}(f,\theta_l,\theta_m)$ is the spectra on-source at the position
$(\theta_l,\theta_m)$, and $\emr{REF}(f,\theta_{l0},\theta_{m0})$ is a
reference spectrum observed at a fixed position $(\theta_{l0},\theta_{m0})$
of the sky where the source does not emit. This reference spectrum is used
1) to correct for the shape of the frequency bandpass, and 2) to subtract
the contribution of the atmosphere to the measured signal. The RMS noise
level will be directly proportional to the system temperature that is the
calibration factor needed to get the right intensity units. Using
  the same reference spectrum for several adjacent pixels introduces a
  slight spatial correlation in the noise properties.
  Section~\ref{sec:noise:spatial:power:density} characterizes this in
  detail.

\subsection{Spectral resampling and spatial gridding}

We wish to study the variations of the emission of a given line as a
function of the position on the sky. We thus need to obtain a
position-position-frequency cube centered around the line rest frequency in
the source rest frame (see Table~\ref{tab:lines}), which is tagged by the
typical velocity of the source in the LSRK frame. However, the gas in a
molecular cloud experiences turbulent motions. These hypersonic motions
imply a combination of a broadening of the linewidth compared to the
natural thermal linewidth and a shift in frequency of the line peak due to
the Doppler effect associated with the large scale velocity gradients. Both
effects are used to probe the kinematics of the molecular gas where star
forms~\citep[see, \eg,][]{orkisz2017,orkisz2019,gaudel2022gas}.

In order to study the kinematics of the gas traced by different molecules,
it is easier to compare spectral line cubes that share the same spatial and
velocity grid. Appendix~\ref{app:doppler} describes the impact of the
Doppler effect on radio-astronomy line cubes. The velocity axis is linked
to the frequency axis through Eq.~\ref{eq:doppler:radio}. In particular,
the velocity resolution associated for a given line is inversely
proportional to the line rest frequency for a spectrum regularly sampled in
frequency. Getting the same velocity axis for the different tracers around
their rest frequencies requires resampling the spectra in velocity. We
choose to resample all the spectra to 0.5\kms{}, which corresponds to the
spectrometer velocity channel spacing at the highest observed frequency in
our data, \ie, the frequency of the \twCO{} \Jone{} line. This means that
all other spectral line cubes will be oversampled along the spectral
axis. As the imperfect Doppler tracking also implies a resampling of the
spectral axis, we correct for both effects in a single resampling
step. This resampling is done by simple linear split (or integration) of
the adjacent channels when the target spectral resolution is narrower (or
respectively wider) than the original one. This ensures that the line flux
is conserved.

\FigDataImages{} %
\FigDataSpectra{} %

At this point, the data are thus a set of spectra regularly sampled on the
same velocity grid. They are also regularly sampled spatially but with
small spatial shifts between two rows along the scanned direction because
the data acquisition only starts when the telescope scanning velocity is
constant, and this event has a relatively uncertain position on the sky for
each line.  We thus need to ``grid'' the spectra on a regular spatial
grid. This is done through a convolution with a Gaussian kernel of full
width at half maximum $\sim1/3$ of the IRAM 30m telescope beamwidth at the
considered rest line frequency.  This operation conserves the flux and
degrades the telescope point spread function width by $\sim 9\%$. Here
again we choose the same spatial grid for all the lines. We set the pixel
size of $9''$ in order to comply with the Nyquist criterion for the studied
line that has the highest frequency. The other spectral line cubes will be
spatially oversampled.

We now end up with one position-position-velocity cube per studied
line. Each cube contains 240 velocity channels times $1074 \times 758$
pixels. The size of the voxels are $9''\times 9''\times 0.5\kms$. The
velocity axis is centered around the rest frequency of the associated
line. While the spatial and spectral grid are common to all cubes, the
spatial and spectral response inversely scales as the line rest
frequency. To ease the computation of line ratios, the cubes are often
convolved with a Gaussian kernel to reach the same angular resolution as
the telescope response of the line that has the smallest rest
frequency. This is the case for the cubes provided in the first public data
release of the ORION-B project\footnote{It is available on the IRAM large
  program archive at \url{https://oms.iram.fr/?dms=frontpage}.}, where the
provided cubes are smoothed to a common resolution of $31''$. In contrast,
no action is in general taken to get a common spectral resolution because a
large fraction of the analysis just relies on the intensity integrated on
the full line profile.


\section{Properties of the signal in two ORION-B spectral line cubes}
\label{sec:signal}

\FigDataHisto{} %

We here analyze the signal properties of two radio-astronomy line cubes
from the ORION-B dataset (namely, the \thCO{} \J10{} and \CseO{} \J10{}
cubes\footnote{These cubes are available on the ORION-B project web page at
  \url{https://www.iram.fr/~pety/ORION-B/data.html}.}). This analysis will
lay out the ground for the innovations proposed in
Sect.~\ref{sec:autoencoder:optimized}.

\subsection{Spatial and spectral means}

A spectral cube contains two spatial dimensions and a spectral
dimension. Figure~\ref{fig:data:images} compares the map of the emission
averaged over the spectral axis for the two cubes. The most obvious
differences are the intensity dynamics (defined as the ratio of the cube
peak intensity to the typical noise level) and the signal-to-noise ratios.
The \thCO{} \Jone{} mean emission has an intensity dynamic of at least a
factor 10. But a fraction of the voxels of the \mbox{\thCO{} \Jone{}} cube still
lies at signal-to-noise ratio lower than 5. The \CseO{} \Jone{} mean
emission mostly looks like noise. Only an astronomer knowing the shape of
the source may guess the existence of some signal on the southeastern part
of the image near NGC\,2023 and NGC\,2024.

Figure~\ref{fig:data:spectra} compares the spectra averaged over the
observed field of view, as well as the minimum and maximum spectra for the
two cubes. The line signal is sparse along the spectral axis: The mean
spectra of the line cubes show signal only between about $-0.5$ and
$16.0\kms$, \ie, a small fraction of the measured channels.  These spectra
confirm the difference already seen for the intensity dynamics and
signal-to-noise ratios. The sparsity of the line signal along the spectral
axis allows us to estimate the noise level. Assuming that the noise follows
a centered Gaussian distribution of RMS $\sigma$, the difference between
the minimum and maximum spectra is $6\sigma$ for $99.7\%$ of the
samples. This gives a typical noise level of about 0.1\,K in our case. The
dynamical range of the line cubes are thus of the order of 430 and 20 for
the \thCO{} \Jone{} and \CseO{} \Jone{} lines, respectively. The spectra in
the \CseO{} \Jone{} cube must be spatially averaged in order to clearly
detect a mean spectrum because the typical signal-to-noise ratio of this
cube is of the order of 1.

\subsection{Histograms of the measured intensities}

Figure~\ref{fig:data:histo} compares the histograms of the intensities for
the two cubes. On each panel, three noise histograms are displayed: The
black one uses all the channels, while the green and red ones use the
channel with mostly signal or noise, respectively. The left column shows
the histograms over the full interval of intensities. These ``signal''
histograms show that the bright end of the \mbox{\thCO{} \Jone{}} intensities
follow an exponential distribution. The right column zooms in over the
faint intensity edge of the histogram. These two ``noise'' histograms are
close to a Gaussian distribution. They are centered on zero by construction
because of the baseline removal.

\subsection{Signal redundancy among the channels}
\label{sec:channel:correlations}

\FigDataSelectedChannels{} %

Figure~\ref{fig:data:selected:channels} compares the spatial distribution
of the signal for two channels of the \thCO{} \Jone{} cube. The two chosen
channels are displayed as the red vertical lines in
Fig.~\ref{fig:data:spectra}.  They are centered on the two main velocity
components of the Orion B molecular cloud~\citep{pety2017}. These channels
display different spatial patterns and are thus quasi-independent, \ie, the
knowledge of the first pattern provides no information on the shape of the
second pattern.

To better quantify this phenomenon, we compute the Pearson correlation
coefficient and the mutual information between each pair of channels. The
former highlights linear relationships between two channels while the
latter is able to capture both linear and nonlinear relationships.  The
absence of a linear correlation does not mean either independence or the
absence of redundancy to be exploited for information extraction.  The
computation of the mutual information is thus desirable because, as shown
by~\citet{licciardi2018}, the relations between the channels of
hyperspectral cubes are sometimes strongly nonlinear.  It quantifies
whether one can predict one quantity knowing the other one, even though the
relationship is nonlinear. It is equal to 0 if and only if both variables
are statistically independent. More details are given in
Appendix~\ref{app:mutual:information}. The mutual information is
numerically computed by approximating the joint distribution with nearest
neighbors~\citep{kraskov2004}. In order to have homogeneous and comparable
results, we express the correlation coefficient in bits of information as
the mutual information \citep{gelfand1959calculation}. If $\rho(X,Y)$ is
the Pearson correlation coefficient between $X$ and $Y$, it can be
expressed in bits of information through
$\mi=-0.5\,\log_2\bracket{1-\rho(X,Y)^2}$. This quantity diverges when the
relationship between the two variables is deterministic. We thus blank the
diagonal coefficients.

The top panel of Fig.~\ref{fig:data:corr:mi} shows the linear relation
between two channels. The linear correlation of the \thCO{} \Jone{} cube
has significant values only in two regions: 1) along the diagonal because
the spectral response of the radio-astronomy spectrometer is slightly
larger than one channel (see Sect.~\ref{sec:noise:spectral:power:density}),
and 2) for the $[3,15\kms]$ velocity range, where the signal sits.  The
bottom panel of Fig.~\ref{fig:data:corr:mi} shows the image of mutual
information that quantifies any relation. Large values of the mutual
information gather into two main groups related to the two velocity
components of the Orion B cloud at $6\kms$ and $11\kms$. Moreover, there is
a faint correlation between the two main velocity ranges. In the signal
region, the coefficient values fall by a factor of $\sim 10$ at a typical distance of 3 or 4 channels. We will call this distance mutual information scale in Sect.~\ref{sec:intrinsic:dimensions:results}. In other words, the mutual information scale is small for the \thCO{} \Jone{} cube.

\FigDataCorrMI{} %


\section{Noise properties}
\label{sec:noise}

\FigDataNoiseDistribution{} %

We next characterize the noise properties inside the acquired
radio-astronomical cubes. In particular we compute the noise spatial and
spectral power density.\footnote{To be precise, we could use the complete
  formulation, \ie, noise spatial and spectral power spectral density. This
  however introduces a confusion between the ``spectral'' (frequency,
  wavelength, or velocity) axis of astronomy cubes and the ``spectral''
  density that refers to computations in the Fourier plane. We thus choose
  to remove spectral in ``power spectral density''.}

\subsection{Spatial and spectral levels}
\label{sec:noise:levels}

\FigDataNoiseSpatialPowerDensity{} %

To estimate the noise levels, we assume that the spatial and spectral
variations of the noise are independent of each other, as proposed
by~\citet{leroy2021}.  The noise RMS can then be factored as
\begin{equation}
  \label{eq:noise:factorization}
  \sigma(i_x,i_y,i_c) = \sigma_\emr{spe}(i_x,i_y)\,.\,\sigma_\emr{spa}(i_c),
\end{equation}
where $\sigma_\emr{spe}(i_x,i_y)$ and $\sigma_\emr{spa}(i_c)$ represent the
spatial and spectral variation of the noise RMS computed along the spectral
and spatial axes, respectively. We start by computing the noise RMS of the
channels for each pixel on channels that are devoid of signal.  We then
divide the signal cube by the spatial variations of the spectral RMS,
$\sigma_\emr{spe}(i_x,i_y)$, and we compute the RMS per channel after
masking regions where signal is detected (see
Sect.~\ref{sec:detection}). Moreover, we compute the standard deviation of
the RMS as $\sigma/\sqrt{2s}$ where $s$ is the number of samples used.

The top panel of Fig.~\ref{fig:data:noise:dist} shows the map of the noise
spectral RMS, normalized by its median value, for the \CseO{} \Jone{}
cube. We do not show the result for the \thCO{} \Jone{} cube because it is
similar to the result for the \CseO{} \Jone{} cube. The noise map has an
obvious inhomogeneous spatial distribution with mostly vertical stripes
organized in squares. This reflects the acquisition scheme, where a single
pixel detector is scanned along vertical lines of size of $\sim 1000''$
inside squares. The noise pattern evolves from left to right because the
scanning strategy was optimized during the acquisition of the ORION-B large
program data. For instance, in the middle of the acquisition we tried to
organize the approximately $1000''$-long scans into long vertical lines instead of
squares. However, this increased the striping in the signal images. We thus
decided to come back to an acquisition in consecutive squares to ensure a
better continuity of the signal.

The noise comes mostly from the atmosphere contribution to the measured
power in radio-astronomy (see Appendix~\ref{app:linearisation}). This
implies that the noise level follows to first order the quality of the
weather. A dry atmosphere during winter observations improve the noise
level by a typical factor of approximately $1.5$ over summer observations for the
two studied lines. This is the origin of the large variations of the noise
level from one square to another. The amount of atmosphere that emits
depends on the source elevation. It is minimum at zenith and maximum when
the source rises and sets. Thus, the noise level also follows the elevation
of the telescope at constant weather, and this is the main origin of the
noise level regular variations inside each square.

The bottom panel of Fig.~\ref{fig:data:noise:dist} shows the variations of
the spatial RMS of the noise with the velocity.  The line cubes show
spectral variations of the noise between $-2$ and $+4\%$ with two
characteristic patterns. First, there is an oscillating pattern that
directly comes from the resampling of the spectra along the spectral
axis. Superimposed, there is also an increase of the noise level of about
$2\%$ following more or less a boxcar function between $-10$ and
$+30\kms$. This is related to the baseline removal step during the
reduction. This step is required to remove remaining atmospheric residual
signal after the atmosphere calibration. It is done by fitting a Chebyshev
polynomial of low order outside the velocity window where the signal
appears with some margin to avoid biasing the baseline by signal at low
signal-to-noise ratio in the line wings. The baseline substracted inside
the signal window is then interpolated using the fitted Chebyshev
coefficients. We here used a polynomial order of degree 1 outside the
$[-10,+30\kms]$ signal window.

\subsection{Noise spatial power density}
\label{sec:noise:spatial:power:density}

We first compute the spatial 2D Fourier transform of the \mbox{\CseO{}
  \Jone{}} cube for 90 channels devoid of signal, from $-50$ to
$-5\kms$. We then compute the square of the modulus of the Fourier
transform, and we finally average the 90 resulting images. This gives an
estimation of the noise spatial power density.

We use the radio-astronomy convention to define the conjugate coordinates
of the angular coordinates $(\theta_l,\theta_m)$ relative to the projection
center of the image as $(u,v)$ with
\begin{equation}
  u\,\theta_l= \lambda,
  \quad \mbox{and} \quad
  v\,\theta_m = \lambda,
\end{equation}
where $\lambda$ is the wavelength of the observed line. In our case,
\mbox{$\lambda = 2.67\,$mm}. The conjugate planes are called image and $uv$
planes, respectively. The $(\theta_l,\theta_m)$ and $(u,v)$ coordinates are
expressed in radian and meter, respectively.

The first column of Fig.~\ref{fig:data:noise:spatial:power:density} shows
the obtained noise spatial power density. For a perfect measurement, we
expect to recover an image proportional to $\modulus{\ft{B}}^2$, that is
the square of the modulus of the Fourier transform of the point spread
function of the telescope $B$.  While $\modulus{\ft{B}}^2$ should show a
radial symmetry to first order, we obtain a spatial power density that is
dominated by a structure elongated along the $u$ axis. This structure comes
from correlations in the observed noise between all the spectra belonging
to the same subscan (scanned vertically in this case).

Appendix~\ref{app:nspd} shows that the noise spatial power density is to
first order equal to
$\PSD(u,v) \simeq \PSD_\emr{on}(u,v) + \PSD_\emr{ref}(u,v)$, with
\begin{equation}
  \PSD_\emr{on}(u,v)
  = \Area{pix}\,\paren{\frac{\sigma_\emr{on}}{\sigma}}^2\,\modulus{\ft{B}}^2(u,v),
\end{equation}
and
\begin{equation}
  \PSD_\emr{ref}(u,v)
  = \Area{rect}\,\paren{\frac{\sigma_\emr{ref}}{\sigma}}^2\,\bracket{\sinC{\frac{\Delta\theta_l\,u}{\lambda}}\,\sinC{\frac{\Delta\theta_m\,v}{\lambda}}}^2.
\end{equation}
In these equations, $\Area{pix}$ and
$\Area{rect} = \Delta\theta_l\Delta\theta_m$ are the respective areas of
the image pixel and of any rectangle that shares the same reference
measurement. Moreover,
$\sigma = \sqrt{\sigma_\emr{on}^2+\sigma_\emr{ref}^2}$, and
$\sigma_\emr{on}$ and $\sigma_\emr{ref}$ are the typical standard deviation
of the noise on source and on reference, respectively.

The second and third columns of
Fig.~\ref{fig:data:noise:spatial:power:density} show the resulting model,
and the ratio of the measured and modeled noise spatial power density in
logarithmic scale. In the studied case, the modeling holds for most of the
$uv$ plane.

\subsection{Noise spectral power density}
\label{sec:noise:spectral:power:density}

\FigDataNoiseSpectralPowerDensity{} %

Figure~\ref{fig:spectral_power_density} shows the noise spectral power
density and the noise autocorrelation. To get them, we first compute the 1D
Fourier transform along the frequency axis for the same subcube devoid of
signal. We then compute the square of the modulus of the Fourier transform,
and we average results over the pixels. The autocorrelation function of the
noise is estimated by calculating the inverse Fourier transform of the
spectral power density.

The autocorrelation shows that the correlation between two channels $x[f]$
and $x[f+\delta f]$ becomes zero when
\mbox{$|\delta f| > 2 \times 183.80\kHz$}. This fact leads us to model the
noise spectral autocorrelation with the autocorrelation of a symmetric
finite impulse response filter of the form \mbox{$h = [a\;b\;a]$}, with the
constraint \mbox{$2a^2+b^2 = 1$} in order to preserve the signal power. The
curve on Fig.~\ref{fig:spectral_power_density} shows five nonzero values
because it corresponds to the autocorrelation of the filter. We estimate
\mbox{$h = [0.18\;0.97\;0.18]$} for the \mbox{\thCO{} \Jone{}} and
\mbox{\CseO{} \Jone{}} spectral cubes. The good fit of the noise
autocorrelation with the autocorrelation of this filter indicates that the
noises of pair of channels separated by more than two channels are
uncorrelated. The estimated filter can be used to simulate noise with a
similar spectral power density.

\subsection{Noise PDFs at low and large signal-to-noise ratios}
\label{sec:noise:pdfs}

The measured intensity at pixel $ij$ and velocity channel $c$ is given by
\begin{equation}
  \label{eq:}
  \Int{ijc} = \paren{1+\epsilon_{ij}}\,\bracket{\Sig{ijc} + \Noi{ijc}},
\end{equation}
where \Sig{ijc} is the signal from the source, \Noi{ijc} the additive noise
coming mostly from the atmosphere and the receiver, and $\epsilon_{ij}$ the
relative uncertainty on the calibration gain. We assume that
$\epsilon_{ij}$ is mostly constant over the narrow-band spectra used
here. The values of \Noi{ijc} and $\epsilon_{ij}$ are drawn from two
centered normal distributions of standard deviation $\sigma_{ijc}$ and
$\Sigma$, respectively. Depending on the observed atmospheric window (3 or
1\,mm), the values of $\Sigma$ range from $0.05$ to $0.1$, \ie,
$\epsilon_{ij} \ll 1$ (for details, see the
appendix~\ref{app:calibration:uncertainty}). Thus, there are two main
different limiting regimes that depend on the signal-to-noise ratio
\begin{eqnarray*}
  \Int{ijc}      \sim \Sig{ijc} + \Noi{ijc}          & \mbox{when} & \Sig{ijc} \ll \Noi{ij},\\
  \log \Int{ijc} \sim \log \Sig{ijc} + \epsilon_{ij} & \mbox{when} & \Sig{ijc} \gg \Noi{ijc}.
\end{eqnarray*}
At low signal-to-noise ratios, we can neglect the uncertainty of the
calibration, and the additive noise dominates the uncertainty budget. In
contrast, at high signal-to-noise ratio, we can neglect the additive noise,
and the uncertainty budget is dominated by the multiplicative noise with
$\log \paren{1+\epsilon_{ij}} \sim \epsilon_{ij}$.


\section{The autoencoder neural network as a generic method of dimension
  reduction}
\label{sec:autoencoder:neural:networks}

In this section, we introduce a deep learning method called autoencoder
neural network. We present its default architecture and operation. We then
use it to compute the amount of redundancy available in the input
dataset. In the next section, we will taylor it for molecular line cubes
based on the data analysis performed in section~\ref{sec:dataset}.

\subsection{Neural networks}
\label{sec:neural:networks}

Artificial neural networks are a class of statistical machine learning
methods that were originally designed to simulate the behavior of the
brain. Today, they are widely used in data science because they allow to
easily model any nonlinear functions in high dimensional spaces. More
precisely, we use architectures derived from the multilayer
perceptron~\citep{shalev2014understanding}. Multilayer perceptrons are
composed of a succession of matrix products and nonlinear functions called
activation functions. They are interesting because they are universal
approximators of any continuous function when they have at least one hidden
layer and this layer contains enough
neurons~\citep{hornik1989multilayer}. Appendix~\ref{app:perceptron} gives
more details.

The modeling of a nonlinear function by a neural network can be considered
as a global optimization problem that is solved through stochastic gradient
descent. The user specifies a loss function that will constrain the neural
network to select one family of functions adapted to the considered
problem. The only constraint on the loss function is that it must be
derivable with respect to each parameter of the network, in order to be
able to perform their optimization by the stochastic gradient descent
algorithm~\citep{duda1973}.

\subsection{Autoencoder neural network}
\label{sec:autoencoder:standard}

\FigStandardAutoencoder{} %

Figure~\ref{fig:standard:autoencoder} shows the architecture of an
autoencoder neural network. As the autoencoder described in
Sect.~\ref{sec:generic:denoising}, it is composed of two cascaded parts,
the encoder and the decoder functions that are implemented as two neural
networks. The encoder aims at computing a simplified representation of the
data. The decoder aims at reconstructing the input data as faithfully as
possible from the simplified representation. In our cases, we will choose
symmetrical architectures for the encoder and decoder parts. Nevertheless,
it does not mean that the functions $\mathcal{E}$ and $\mathcal{D}$ are
inverse from each other, as explained in Sect.~\ref{sec:generic:denoising}.

The reduction of dimension space enforced by the autoencoder can be
interpreted as an approximation of a nonlinear principal component
analysis~\citep{licciardi2015}. In the case of noisy data containing signal
with a low dimension representation, this compression should retain the
signal features and filter the noise. As an autoencoder neural network is
designed to identify a low dimension representation of the signal, it
allows one to perform a generic denoising operation. In particular, it
generalizes the denoising operation that can be performed with a principal
component analysis in the case where the signal features are nonlinearly
correlated.

\subsection{Estimating the intrinsic dimension of a dataset}
\label{sec:intrinsic:dimensions}

When denoising by reduction dimension, the amount of denoising is related
to the redundancy in the input data, which allows one to reduce the
dimension without loosing relevant information.  If the dimension of the
input data is called the extrinsic dimension and the dimension of the
bottleneck the intrinsic dimension, we thus wish to measure the intrinsic
dimension of the data. The extrinsic dimension is necessarily greater than
or equal to the intrinsic dimension.

An autoencoder neural network is interesting here because it is a practical
algorithm that encompasses the whole category of methods that assumes a
reduction of dimension to denoise the data (see
Sect.~\ref{sec:introduction}). We use the autoencoder to analyze the
intrinsic dimension of the signal with respect to the extrinsic dimension
of the data, and thus emphasize the amount of redundancy that could be used
to increase the signal-to-noise ratio.

\subsection{Implementation}

\FigDataIntrinsicDimensions{} %
\FigNormes{} %
\FigOptimizedAutoencoder{} %

We define a set of autoencoders whose bottleneck size varies between one
and the extrinsic dimension of the data $(m)$. The loss function is then
minimized for each of these autoencoders.
Figure~\ref{fig:data:intrinsic:dimensions} shows the mean absolute
deviation between the input data and the denoised data as a function of the
bottleneck size $(l)$. The intrinsic dimension is the smallest dimension of
the bottleneck that allows us to reconstruct the signal without significant
loss of relevant information. Two regimes are expected for this curve: A
quick decrease of the mean absolute deviation as long as increasing the
bottleneck size adds useful information to reconstruct the signal, followed
by a constant value of the mean absolute deviation when further increasing
the bottleneck size starts to reconstruct the noise. The threshold between
these two regimes is interpreted as the intrinsic dimension of the
data. This method is directly inspired by the ``elbow method'' used when
denoising with a principal component analysis~\citep{ferre1995selection}.

The choice of the loss function is key to ensure a proper estimation of the
intrinsic dimension. A desirable property is to select encoders that will
maintain independent input variables as independent bottleneck neurons
instead of encoding them as linear combinations.  Using the mean absolute
deviation instead of the more usual mean squared error allows one to avoid
mixing independant inputs. Indeed, let's assume that the data is composed
of two uncorrelated non-Gaussian (\eg, Laplacian) variables of mean 0 and
variance 1.  The encoding of this pair of variables with a single component
(\ie, an autoencoder with a single bottleneck neuron) consists in searching
for the direction that maximizes the norm of the projection in one
direction. As illustrated in Fig.~\ref{fig:normes}, the $L_2$ norm is
invariant to rotation, implying that the maximization of the projection is
not sensitive to rotation and the encoder will mix the two components. In
contrast, the values of the $L_1$ norm varies under rotation, and the
autoencoder will thus avoids mixing the independent pair of variables.  In
other words, if we try to encode the two independent variables with a
bottleneck made of a single neuron, the MSE loss function will constrain
the autoencoder to pay attention to the largest values of the two random
variables and to combine them linearly in order to minimize its value. In
contrast, the mean absolute deviation will enforce a solution where only
one of the two independent variables is encoded in the bottleneck, the
other one being ignored.

\subsection{Comparison of the intrinsic dimension between the ORION-B cubes
  and a typical hyperspectral cube}
\label{sec:intrinsic:dimensions:results}

Figure~\ref{fig:data:intrinsic:dimensions} compares the evolution of the
mean absolute deviation as a function of the dimension of the bottleneck
for two datasets: The ORION-B \thCO{} \Jone{} line cube on the left panel,
and a Earth remote sensing hyperspectral cube, named Indian
Pines\footnote{The latter dataset, named Indian Pines, has been acquired
  with the Airborne Visible/Infrared Imaging Spectrometer (AVIRIS) sensor
  over an agricultural area located at northwestern Indiana, USA. This cube
  is composed of 220 spectral channels ranging from 400\,nm to
  2500\,nm. Its spatial linear resolution is $20\times 20\,$m. It is
  publicly available here
  \url{https://www.ehu.eus/ccwintco/index.php/Hyperspectral_Remote_Sensing_Scenes}.},
that is used to benchmark denoising algorithms on the right panel. This
comparison is useful because \citet{licciardi2018} showed that dimension
reduction with a neural autoencoder is particularly efficient to denoise
the latter dataset.
  
The intrinsic dimension of Indian Pines can be estimated at around 4.  In
constrast, the curve for \thCO{} \Jone{} only has a clear elbow at about
27. This implies that the intrinsic dimension of the signal is close to its
extrinsic dimension. This confirms our previous finding that the measured
mutual information scale is small for the ORION-B line data (see
Sect.~\ref{sec:channel:correlations}).

Two main properties explain the different behaviors of the \thCO{} \Jone{}
and Indian Pine cubes. The astronomy line cube contains many signal-less
channels that are irrelevant for scientific purpose but can be used to
characterize the noise properties. Moreover, the achieved spectral
resolution still limits the amount of redundancy inside the sampled line
profile. In contrast, almost all the channels of Indian Pine cube are
scientifically relevant and (anti-)correlated.  In this respect, denoising
by dimension reduction would be easier for astronomy hyperspectral cubes
observed with direct detection imaging spectrometers used to study the
spectral energy distribution of the sources, including the continuum and
and low to medium resolution spectral line emission, such as the
SPIRE and PACS spectrometers on-board Herschel~\citep{pilbratt2010} or the
MIRI and NIRSpec instruments on-board JWST~\citep{rigby2022}, because such
instruments provide hyperspectral cubes with scientifically relevant
information for each spectral channel.


\section{A locally connected autoencoder with prior information to denoise
  line data}
\label{sec:autoencoder:optimized}

As discussed in the previous section, the reduction dimension of the
ORION-B line cubes is more difficult than in the case of Earth remote
sensing cubes. It is thus all the more important to optimize the structure
of the used autoencoder neural network with sound assumptions to help it
converge on the correct solution. In this section, we propose an innovative
autoencoder structure adapted to the properties of the line cubes. We will
first describe the geometry of the autoencoder that takes into account the
fact that the mutual information scale is small compared to the extrinsic
dimension of the data. We will then propose a loss function that ensures
that channels without signal are set to zero instead of some arbitrary
(small) value.

\subsection{Locally connected autoencoder}
\label{sec:autoencoder:locally:connected}

A typical autoencoder is composed of fully connected layers, \ie, all the
input neurons of the layer are connected to each output neuron (see
Fig.~\ref{fig:standard:autoencoder}). This ensures that all potential
correlations between the input data are explored. In line cubes, only
channels at nearby frequencies are correlated. This means that an
autoencoder would try to learn the numerous combinations of uncorrelated
channels. Figure~\ref{fig:optimized:autoencoder} shows an architecture
where a set of multilayer perceptrons connects adjacent input neurons to
adjacent bottleneck and output neurons. In our case, this means that only
adjacent channels will be autoencoded together. This change introduces a
major difference compared to a typical autoencoder. The latter would
deliver the same result (within numerical approximations) whatever the
ordering of the input neurons. In contrast, our taylored autoencoder
assumes that adjacent channels are linked together. This means that we
introduce the notion of proximity in frequency of the channels inside the
autoencoder architecture.

As a comparison, a convolutional layer\footnote{Unlike a dense layer used
  in a perceptron which is composed of a matrix product, a convolutional
  layer is composed of a linear filter.}~\citep{o2015introduction} would in
addition take into account the order of the channels. However, the applied
convolution filter would be identical for all observed spectra. In other
words, a convolutional layer assumes spectral translation invariance with
respect to the observed spectra while the proposed architecture does
not. In particular, the fact that the signal-to-noise ratio and the amount
of signal information can vary significantly with the frequency would be
ignored with a convolutional layer.
  
For simplicity, we choose a symmetric autoencoder which has a total of four
hyperparameters that must be chosen: 1) $l$, the size of the bottleneck
layer; 2) $p$, the size of the sliding window that connects nearby
channels; 3) $q$, the size of each perceptron layer; and 4) $h$, the number
of hidden layers of each perceptron. We have the following relations:
$l < m$ and \mbox{$p < q < m,$} where $m$ is the number of input and output
channels in the spectrum. The hyperparameters of the taylored autoencoder
may depend on the studied line. For instance, it is likely that the model
for a line such as \thCO{} \Jone{} is more complex than for the
\mbox{\CseO{} \Jone{}} line, implying larger values for $l$ and $h$. The
data analysis performed in Sect.~\ref{sec:channel:correlations} imposes
some constraints. If $r$ is the mutual information scale in channel units,
the optimal window size is \mbox{$p = 2r+1$}. Moreover, $\frac{m}{r}$ is a
(potentially optimistic) lower bound for the size of the bottleneck because
it represents the number of groups of channels that are decorrelated from
each other.

In practice, the simplest implementation of our taylored autoencoder is to
perform a matrix product for each window.  However, the autoencoder will
then perform a large number of consecutive matrix products leading to large
overheads. We instead choose to encode the set of locally connected
perceptrons as a unique fully connected perceptron, where the superfluous
weights are set to 0 during the initialization and the associated gradients
are multiplied by 0 during the training. This requires a single (taylored)
matrix multiplication per layer. The number of free (\ie{} nonzero)
parameters in this optimized autoencoder can be computed directly from the
Python implementation that is available on the project GitHub
repository. In our application, the number of free parameters is only 6\%
of the total number of matrix elements. This eases the training of the
optimized autoencoder.

\subsection{Adding prior information to the optimization problem}
\label{sec:problem_with_prior}

As described in Sect.~\ref{sec:generic:denoising:practice}, denoising by
dimension reduction is an optimization problem that tries to find the
autoencoding function $\cal A$ that will minimize the distance between the
data and its autoencoding, averaged over all the data samples: see
Eq.~\ref{eq:optimization:1} and~\ref{eq:loss:1}. The presence of noise
implies three adaptations of the autoencoder about the definition of its
training loss function. The first one will take into account the important
variation of the SNR (from $<1$ to a few $100$) in radio-astronomy
data. The second one will address the potential unbalance between the
number of voxels that only contain noise and the number of voxels that
actually contain relevant signal. The third one will ensure that the
autoencoder attributes a zero-valued intensity (instead of any other
randomly chosen systematic value) for voxels that only contain noise.

\begin{description}
\item[\textbf{To handle varying SNR values}] the distance is usually
  weighted by the standard deviation of the noise. In our case, the
  baseline part of the spectrum enables us to easily estimate the noise
  standard deviation $\sigma_k$ for the spectrum $d_k$ at pixel
  \mbox{$k = i_x + n_x\,(i_y-1)$}. We thus will modify the loss function as
  \begin{equation}
    \label{eq:loss:2}
    {\cal L}({\cal A},d) = \frac{1}{K} \, \sum_{k=1}^K \paren{\frac{{\cal A}(d_k)-d_k}{\sigma_k}}^2.
  \end{equation}
  This normalization avoids the variation of the data ``energy'' just
  caused by noise, which would overweight the noisiest pixels. We recognize
  here the reduced $\chi$-squared merit function that is regularly used in
  astronomy. In contrast, the machine learning community mostly uses the
  MSE.
\item[\textbf{To address the problem of sparsity of the signal inside the}]
  \textbf{cube,} we will balance the loss function by giving it \textit{a
    priori} information about the channels that have a large probability to
  be just noise. To do this, we first segment the
  position-position-frequency cube into signal and noise samples (see
  Sect.~\ref{sec:detection}). We then modify the loss function as
  \begin{equation}
    \label{eq:loss:3}
    {\cal L}({\cal A},d) = \frac{1}{K}\,\sum_{k=1}^K \cbrace{%
      \begin{array}{l}
        \frac{1}{\sum_{j=1}^J w_{jk}}
        \sum_{j=1}^J w_{jk} \paren{\frac{{\cal A}(d_{jk})-d_{jk}}{\sigma_k}}^2 \\
        + \\
        \frac{1}{\sum_{j=1}^J (1-w_{jk})}
        \sum_{j=1}^J \paren{1-w_{jk}} \modulus{\frac{{\cal A}(d_{jk})-0}{\sigma_k}}^q
      \end{array}
    }
  \end{equation}
  where $w_{jk}=1$ for a channel $j$ of spectrum $k$ dominated by signal,
  and $w_{jk}=0$, elsewhere.  The normalization factors ensure that
  noise-only \mbox{($\emr{SNR} < 1$)} samples do not dominate the loss
  function. This solves the potential unbalance between signal and noise
  samples inside each spectrum. While the architecture of the optimized
  autoencoder does not use the spatial information, the segmentation used
  in the proposed loss function introduces some spatial information as it
  is a method that works in the position-position-frequency space.
\item[\textbf{To ensure that noise-only samples deliver 0}] instead of a
  small random value, we use the $L_q$ norm\footnote{The $L_q$ norm of the
    vector $\mathbf{x} = (x_1, ... , x_n)$ is defined as
    \mbox{$\norm{q}{\mathbf{x}} = \paren{\sum_{i=1}^{n}
      \modulus{x_i}^q}^{1/q}.$}}, with \mbox{$q\in ]0,2]$}, for samples that
  are mostly noise. This enforces the training to choose either 0 or the
  autoencoded (denoised) value of the data, \ie, \mbox{${\cal A}(d_{jk})$}.
  The denoised value of the data will be selected when the data sample has
  a statistical signature too far from random Gaussian noise. The
  hyperparameter $q$ allows one to finely control the asymptotic behavior
  of the penalty of voxels containing only noise: The closer $q$ is to 0,
  the larger the penalty applied to an autoencoded value close to zero. In
  this study, we chose $q=1$.
\end{description}

\subsection{Detecting significant signal}
\label{sec:detection}

\FigDetectionSNR{}%
\FigDetectionSegments{}%
\FigDetectionMoments{}%

The \CseO{} \Jone{} is characterized by a low signal-to-noise ratio. The
best way to detect signal in such a condition is to correlate the noisy
measurement with the expected shape of the signal and to threshold the
output because the probability that random noise reproduces the expected
shape is negligible.  This technique, named matched filtering, is all the
more effective when the shape of the signal is accurately known. For
example, if one aims at detecting a point source, we just need to know the
point spread function of the instrument. Correlating the noisy measurement
with the point spread function thus not only delivers an optimal way to
detect point sources, but it also improves the detection of spatially
resolved sources. Indeed, adjacent pixels can be thought as measurements of
the same source where the noise is uncorrelated from one pixel to
another. As the pixel size is chosen to at least Nyquist-sample the point
spread function, any source will be spread over at least four contiguous
pixels, and the signal-to-noise ratio after correlating with the point
spread function will be much higher than the signal-to-noise ratio per
pixel of the original image. As this makes no assumption on the shape of
the source, this is a simple way to optimize the detection of any kind of a
resolved source. In summary, while matched filtering is the optimal way to
detect point sources, it also improves the detection of resolved sources
because it smoothes the data to an angular resolution larger by $\sqrt{2}$
and thus naturally increases the signal-to-noise ratio per pixel.

Figure~\ref{fig:detection:SNR} shows the map of the maximum and minimum
signal-to-noise ratio per spectrum before and after correlation of the
\CseO{} \Jone{} line cube by the telescope point spread function. In both
cases, the signal-to-noise ratio is defined as
\begin{equation}
  \emr{SNR}(i_x,i_y,i_c) = \frac{d(i_x,i_y,i_c)}{\sigma(i_x,i_y,i_c)},
\end{equation}
where $d(i_x,i_y,i_c)$ and $\sigma(i_x,i_y,i_c)$ are the
position-position-velocity cubes of intensities and noise RMS,
respectively. The computation of the noise RMS is described in
Sect.~\ref{sec:noise:levels}. When correlating the cube by the instrument
response, the maximum of the SNR improves by a factor on the order of $2$ from $9.7$ to
$16.8$, and the percentage of pixels whose maximum SNR value is above 5
increases from $0.13$ to $0.75\%$. In contrast, the minimum SNR value is
relatively stable ($-6.7$ vs $-6.1$) as expected when the noise is (mostly)
uncorrelated between adjacent pixels.

\FigChannelComparison{} %

The SNR cube can then be thresholded to yield a 3D mask of detected
pixels. On one hand, we wish to reduce the number of false positives. This
requires to use a relatively high threshold value. Indeed, for a Gaussian
additive noise, even using a SNR threshold value of 3 yields about 0.3\% of
false positives, \ie, approximately $10^5\,$voxels even when assuming that the
signal can be present only between $-5$ and $20\kms$. On the other hand, we
wish to reduce the number of false negatives. In millimeter
radio-astronomy, a large fraction of the source flux frequently has SNR
values lower than 3. Using a too high SNR threshold value thus implies a
large quantity of false negative pixels.

The first way to improve the tradeoff between the requirements to minimize
the number of false positives and negatives uses again the fact that the
noise distribution is (mostly) uncorrelated between contiguous pixels. It
is indeed possible to segment the cube in regions contiguous in the
position-position-velocity space and for which all pixels have a SNR value
above a given threshold. In practice, we define segments of voxels
  contiguous in the position-position-velociy space, which satisfy the SNR
criterion. When a voxel is added to the current segment, we check
whether the segment should be merged with a segment already
  defined in the previous row of the current image or the previous image of
  the cube. The pixels that do not satisfy the criterion are put in a
specific segment regardless of their position in the
cube. Segmenting in contiguous regions above a given threshold was proposed
by \citet{pety2003} along the spectral axis and \citet{rosolowsky2006} in
3D. When adjacent samples have uncorrelated noise levels, the probability
of a false negative decreases when the total SNR of the region (defined as
the sum of the SNR over all the pixels of the region) increases. Hence
sorting the segmented regions by decreasing total SNR and selecting the
first few ones minimizes the chance 1) to overlook large regions at
relatively low values of the mean SNR, and 2) to yield too many false
positive regions.

Figure~\ref{fig:detection:segments} shows the evolution of three properties
of the 3D segments obtained for the \CseO{} \Jone{} line, and sorted by
decreasing value of the SNR summed over their voxels (hereafter named
segment total SNR). The three properties are the number of voxels inside
each segment, the segment total and mean SNR. These properties are shown
for two different SNR thresholds (1 and 2) used during the cube
segmentation process. Figure~\ref{fig:detection:moments} shows maps of the
\begin{itemize}
\item peak intensity, $\max_{i_c} I(i_c)$;
\item line integrated intensity, $\sum_{i_c} I(i_c) dv$;
\item and centroid velocity,
  \mbox{$\cbrace{\sum_{i_c} v(i_c) I(i_c)} / \cbrace{\sum_{i_c} I(i_c)}$}.
\end{itemize}
We compute them by including the voxels that belong to the first 200
segments. In all generality, the number of segments included is a
compromise between including only the segments with the highest total SNR
and enough segments with a mean SNR larger than 3. Two hundred segments is
a good compromise when the SNR threshold is 2. We here use the same number
of segments when the SNR threshold is 1 in order to make a comparison
without changing too many parameters at a time.

\FigStatisticalComparisonDenoisingRMSE{} %
\FigStatisticalComparisonDenoisingDenoisedVsRaw{} %

For the \CseO{} \Jone{} line, the number of voxels per segment varies from
more than 10 millions to about 1, in comparison with the 195 millions of
voxels present in the cube. The total SNR follows a similar trend because
the mean SNR per voxel is low. In contrast, the mean SNR and images have a
different behavior depending on the SNR threshold.
\begin{description}
\item[\textbf{For a threshold of 1,}] the segment mean SNR is always
  smaller than 3. It is constant at about 1.5 before oscillating. Voxels
  have been selected over almost all the field of view and it is difficult
  to see any structured signal in the three associated maps.
\item[\textbf{For a threshold of 2,}] the segment mean SNR starts to
  decrease or oscillates above 3 before converging to about 2.5 with an
  increasing dispersion. The signal is now pretty well defined in the three
  associated maps, even though some vertical striping is sometimes still
  visible.
\end{description}
These properties can be understood by the fact that for uncorrelated
Gaussian noise, the probability to have the intensity of one of the 6th
closest neighbors to any voxels above 1, 2 or $3\sigma$ is
$0.90 = (1-0.683^6)$, $0.25$, and $0.02$ respectively.  This implies that
any voxel has a large chance to be part of the first segment for an SNR
threshold of 1, a minor chance for a threshold of 2, and a negligible
chance for a threshold of 3.


\section{Denoising performances}
\label{sec:denoising:performances}

We here compare the denoising performances between our taylored autoencoder
and the ROHSA algorithm\footnote{We also compared with the GAUSSPY$+$
  algorithm~\citep{riener19}, which guesses the number of fitted Gaussian
  components per pixels instead of fixing it over the full field of view as
  ROHSA does. While both algorithms deliver solutions with slightly
  different systematic deviations, the differences are not compelling
  enough to warrant presenting both of them.} that we shortly summarized in
Sect.~\ref{sec:generic:denoising:practice}. We do this comparison on the
\mbox{\thCO{} \Jone{}} cube that displays a large SNR range. Our autoencoder
neural network and ROHSA share several properties. They propose a
representation of the data that can be interpreted as denoising by
dimension reduction. They work mainly on individual spectra with a
regularization term that introduces some spatial information about the
data. They nevertheless differ in the family of functions assumed to encode
the data. ROHSA assumes that the signal is composed of a limited number of
Gaussian functions whose amplitude, position, and standard deviation are
spatially regularized. Our autoencoder assumes that the data can be
approximately classified as noise and signal pixels, and that the scale of
mutual information between channels is small compared to the number of
channels in the spectra.

\subsection{Detailed setups of the autoencoder and ROHSA}

We use the Python framework PyTorch to implement our numerical neural
network experiments~\footnote{\url{https://pytorch.org/}}. The segmentation
of the line cubes is implemented in a new IRAM software named CUBE and
distributed inside GILDAS\footnote{The GILDAS software are distributed here
  \url{https://www.iram.fr/IRAMFR/GILDAS/}.}. The associated Python and
CUBE scripts are available in a GitHub
repository\footnote{\url{https://github.com/einigl/line-cubes-denoising}}.

We use the approximately $800\,000$ spectra of 240 channels as input to the
autoencoder. We tagged as mostly signal the voxels that belong to the first
200 segments obtained with a SNR threshold of 2, and the reminders as
mostly noise. The hyperparameters of the autoencoder were optimized as
follows. The width of the sliding window is set at 7 channels according to the mutual information scale (see Sect.~\ref{sec:channel:correlations}). Most
of the other hyperparameters were set with a typical cross validation
procedure~\citep{refaeilzadeh2009cross}. In short, we first defined a set
of possible values to explore. For each set of hyperparameters, we then
optimized the network on a training dataset and we compute its performance
on a different validation dataset. In order to reduce the variability of
the results depending on the choice of the training and validation sets,
this procedure is performed several times, varying the test and validation
sets so that each sample has been selected once in the validation set
during the procedure. This gives for the local encoder: A bottleneck size
of 75\% the number of input channels (here 180), and 3 hidden layers of
size \mbox{$[35, 14, 7]$} per perceptron. During this cross validation
procedure, the hyperparameters that are assumed noncritical are fixed to
usual values: The \texttt{Adam} stochastic optimizer~\citep{kingma2014adam}
was used with a batch size of 100, 50 epochs, and a learning rate that
decreases exponentially from $10^{-3}$ to $10^{-6}$.

Instead of trying to optimize the hyperparameters of ROHSA for denoising,
we used the ones derived by~\citet{gaudel2022gas} when trying to decompose
the spectra into a set of coherent velocity layers in order to study the
velocity field around the filaments of gas where stars will form. The
number of Gaussians was set to 5 for the \thCO{} \Jone{} cube, and the
Lagrangian multipliers used to regularize the maps of Gaussian amplitude,
position, and standard deviation were
\mbox{$\lambda_a = \lambda_\mu = \lambda_\sigma = 100$}.

\subsection{Results}

\FigStatisticalComparisonDenoisingMeanSpectra{} %
\FigStatisticalComparisonDenoisingMomentsSignal{}%
\FigStatisticalComparisonDenoisingMomentsDiff{}%

Figure~\ref{fig:denoising:comparison:channels} compares the raw images with
the denoised ones obtained with the autoencoder and ROHSA for four
different velocity channels that were chosen in the line wings because
denoising of the additive component is expected to act mostly at low to
intermediate SNR. The two algorithms produce similar results to first
order. They both set noise-only voxels to a value close to zero. The shape
of significant signal is kept, and the residuals mostly look like noise. A
closer look suggests that ROHSA delivers signals that are more spatially
coherent than the autoencoder at low SNR but this stays within the noise
level. At intermediate SNR, ROHSA deforms the signal more than the
autoencoder as can be seen in the residuals of the channels at 13.5\kms.

A more quantitative comparison can be seen in
Fig.~\ref{fig:denoising:comparison:rmse} that shows the spatial variations
of the spectral RMS of the residual cubes and their ratio with the spectral
RMS of the raw data. The spatial variations of the spectral RMS show that
both algorithms recover the rectangular pattern coming from the ON-REF
acquisition method. However, a significant part of the signal appears in
the ROHSA residuals, while only a few point sources appear in the
autoencoder residuals. The signal that remains in the autoencoder residuals
is coming from defaults in the signal tagging procedure. The better
preservation of the signal by the autoencoder goes hand in hand with a
slight under-denoising. Indeed, the map of the spectral RMS of the
residuals normalized by the spectral RMS of the noise is on average lower
than 1 in regions that have been tagged as mostly signal. In other words,
the denoised output is closer to the raw input than it should be in case of
perfect denoising. In contrast, the residuals of ROHSA better recover the
noise level at low SNR at the price of more distortion of the signal at
high SNR.

Figure~\ref{fig:denoising:comparison:histo} compares the joint histogram of
the denoised vs the raw intensities. A perfect denoising of the noise
additive component would deliver a joint histogram along the diagonal at
large SNR and an histogram whose dispersion is very asymmetric around zero:
The distribution should have the same dispersion as the noise along the raw
intensity axis and a narrow dispersion along the denoised intensity
axis. The autoencoder succeeds in mimicking the identity function with a
good approximation for signal above $20\sigma$, \ie, a much lower value
than ROHSA. The two algorithms have different behaviors around zero
intensity. On one hand, ROHSA biases the denoising to positive intensities
resulting into a larger vertical size of the histogram, \ie, larger
dispersion along the denoised intensity axis for positive values. On the
other hand, the autoencoder slightly biases the denoising to positive
values for positive raw intensities and to negative values for negative raw
intensities. The bias is more significant for the negative part and can be
tracked in the raw cube to voxels in the surrounding of obviously positive
signal. We interpret this as the consequence of the matched filtering step
that includes in the mostly signal mask negative intensities at the edges
of strong signal.

The denoising quality must also be judged on quantitative estimators that
strongly differ from the loss function.
Figure~\ref{fig:denoising:comparison:spectra} compares the averaged spectra
before and after denoising for the autoencoder and ROHSA. Both the
autoencoder and ROHSA deliver an overall positive bias on spectral regions
that contain the signal but the bias is about twice lower for the
autoencoder. This means that the algorithms slightly bias positively the
total flux of the source. Finally,
Figs.~\ref{fig:denoising:comparison:moments:signal}
and~\ref{fig:denoising:comparison:moments:diff} compare the spatial
variations of the properties of the \thCO{} \Jone{} line before and after
denoising. The results on the raw data cube can be considered as
unbiased. The properties are computed on the raw and denoised data in
exactly the same way. In particular, we used the same spectral window
$[-5,21\kms]$ to compute the line moments. In addition to the peak
intensity, line integrated intensity, and centroid velocity defined in
Sect.~\ref{sec:detection}, we compute
\begin{itemize}
\item the robust line width that is the ratio of the line integrated
  intensity by the peak intensity; this value would be equal to the line
  full width at half maximum for a Gaussian shape;
\item the line velocity dispersion, computed as the square root of
  $\cbrace{\sum_{i_c} \bracket{v(i_c)-C}^2 \,I(i_c)} / \cbrace{\sum_{i_c}
    I(i_c)}$.
\end{itemize} Denoising has a higher impact on the higher order moments of
the line, \ie, the centroid velocity and the velocity dispersion. To first
order, the autoencoder and ROHSA algorithms give similar results. In
particular, the histograms of the residuals between these two methods are
all centered on zero. Moreover, they both set low maximum intensities
closer to zero than the raw data, as expected for a denoising
algorithm. Looking in more detail, differences appear in regions of low to
intermediate SNR. ROHSA better removes the striping pattern of the noise in
regions devoid of signal but it does this by biasing positively the maximum
intensity and the line integrated intensity. The velocity of the maximum is
better preserved by the autoencoder than by ROHSA, but the two algorithms
deliver similar centroid velocity results. Finally, the line width
estimator delivers narrower linewidths on the autoencoder data than on
ROHSA data, in particular in regions of low SNR.

\subsection{Perspectives}
\label{sec:perspectives}

Our autoencoder does not rely on the spatial information, in particular,
the spatial correlations of the noise. Wavelet scattering transforms and
wavelet phase harmonic transforms are recent tools that allow to
characterize the spatial texture of data in statistical ways with only a
few hundred coefficients~\citep{allys2019,levrier2021}. This can be used to
denoise astrophysical data as proposed by~\citet{regaldo2020}.
Investigating whether this would improve the denoising performances
achieved here will be the subject of a forthcoming paper.

\section{Conclusion}
\label{sec:conclusion}

In this paper, we proposed a promising approach to denoise radio-astronomy
line data cubes, inspired by a method developed to denoise hyperspectral
cubes in Earth remote sensing. To do this, we first characterized in-depth
the properties of the noise and signal for two radio-astronomy
position-position-velocity cubes that are part of the ORION-B IRAM 30m
large program, namely the \mbox{\thCO{} \Jone{}} and \mbox{\CseO{} \Jone{}}
cubes.
\begin{itemize}
\item The additive noise is well represented by a Gaussian random
  variable. Its RMS value varies spatially and spectrally. It can be
  modeled as the product of a spatial and a spectral contributions.
\item The spatial variations come from a combination of the source scanning
  strategy, variations of the atmospheric conditions between, \eg, winter
  and summer runs, and the source elevation during each observing session.
\item The spectral variations mostly have two origins. First, the
  resampling (currently) required to correct for Doppler effects in
  wide-bandwidth observations implies a sinusoidal oscillation of the noise
  level with frequency. Second, the interpolation of the polynomial fit of
  the baseline also slightly increases the noise RMS in the line frequency
  range.
\item The noise spatial power distribution can be modeled as the sum of two
  components: 1) the square of the Fourier transform of the telescope point
  spread function, and 2) the modeling of the noise correlation introduced
  by sharing the same reference spectra among many on-source spectra.
\item The noise spectral autocorrelation can be modeled by the
  autocorrelation of a finite impulse response filter of shape
  $[0.18\;0.97\;0.18]$. This implies that the noise between pairs of
  channels is uncorrelated as long as their distance is larger than 2
  channels.
\end{itemize}

Moreover, the signal is sparse along the spectral axis. This allows an easy
estimation of the noise level and the associated SNR. This SNR varies from
less than 1 to several hundred, mostly because of the large intensity
dynamic range. The uncertainty budget is dominated by additive noise at low
SNR, but it becomes dominated by multiplicative noise due to the uncertain
calibration when the SNR is larger than the inverse of the RMS of the
calibration uncertainty: on the order of $20$ in our case. In this paper, we only
denoised the low SNR part of the observations dominated by additive noise.

We then look at the cube as a set of spectra that are individually denoised
by dimension reduction. This method assumes that there is linear or
nonlinear redundancy between the data features (here the channels of any
spectrum). This hypothesis is well verified by standard hyperspectral cubes
usually produced in Earth remote sensing. A mutual information computation
shows that this hypothesis is more problematic for radio-astronomy line
cubes, because the signal information decorrelates quickly from one channel
to another at the obtained spectral resolution. From this viewpoint,
denoising by dimension reduction would be more adapted to astronomy
hyperspectral cubes observed with direct detection imaging spectrometers
used to study the spectral energy distribution of the sources.  When
dealing with cubes that only contain spectrally resolved line emission, any
denoising method by dimension reduction must thus take into account the
fast decorrelation of channels that characterize these cubes.

An autoencoder is a nonlinear low rank deep learning denoising method
whose goal is to minimize the distortion of the signal. We adapted the
typical architecture to our line data as follows.
\begin{enumerate}
\item The proposed architecture takes into account the fast decorrelation
  of the signal as a function of frequency.
\item We take into account the sparsity of the signal inside the spectrum
  by adapting the loss function of the autoencoder depending on whether the
  voxels contain mostly signal or mostly noise. This implies an \textit{a
    priori} position-position-frequency classification algorithm.
\item For ``signal'' voxels, we weight the distance between the data and
  the autoencoded data by the inverse of the noise variance. For ``noise''
  voxels, we use the $L_1$ norm between the autoencoded data and 0 to
  ensure that the autoencoder will not create/destroy flux for low SNR
  voxels.
\end{enumerate}

We finally compare the denoising performance to that achieved by the ROHSA
algorithm that represents the spectra as a set of Gaussian fits.  While
ROHSA allows one to decompose the signal into velocity
layers~\citep[\eg,][]{gaudel2022gas}, the denoising performances of the
proposed autoencoder are higher. The latter allows us to increase the
signal-to-noise ratio (SNR) in pixels with low SNR while preserving the
shape of spectra in high SNR pixels.


\begin{acknowledgements}
  This work is based on observations carried out under project numbers
  019-13, 022-14, 145-14, 122-15, 018-16, and finally the large program
  number 124-16 with the IRAM 30m telescope. IRAM is supported by INSU/CNRS
  (France), MPG (Germany) and IGN (Spain).
  This work was supported by the French Agence Nationale de la Recherche
  through the DAOISM grant ANR-21-CE31-0010, and by the Programme National
  ``Physique et Chimie du Milieu Interstellaire'' (PCMI) of CNRS/INSU with
  INC/INP, co-funded by CEA and CNES.
  This project also received financial support from the CNRS through the
  MITI interdisciplinary programs.
  JRG and MGSM thank the Spanish MCINN for funding support under grant
  PID2019-106110G-100.
  Part of the research was carried out at the Jet Propulsion Laboratory,
  California Institute of Technology, under a contract with the National
  Aeronautics and Space Administration (80NM0018D0004). D.C.L. was
  supported by USRA through a grant for SOFIA Program 09-0015.
  We thank Antoine Marchal and Marc-Antoine Miville-Deschenes for their
  help in using the ROHSA algorithm.
  We thank the referee for valuable comments that helped us to improve the
  manuscript.
\end{acknowledgements}


\bibliographystyle{aa} %
\bibliography{main.bib} %

\begin{appendix} %


\section{Doppler effect and implied spectral resampling}
\label{app:doppler}

The observed lines are emitted in the source frame at the line rest
frequency, \eg, $f^\emr{rest} = 110.20135$\,GHz for the \thCO{} \Jone{}
line. The relative motion between the observatory and the Orion B molecular
cloud in the Milky Way implies that the lines are recorded in the
observatory frame at a frequency shifted by the Doppler
effect. \citet{pety2011} describe in depth the consequences of this effect
on the spectral data. In short, this effect can be approximated to first
order in the Doppler parameter $\frac{v}{c}$ (radio velocity convention) as
\begin{equation}
  \label{eq:doppler:radio}
  \frac{f^\emr{rest}-f^\emr{obs}}{f^\emr{rest}} = \frac{v_\emr{sou/obs}}{c},
\end{equation}
where $c$ is the speed of light, $v_\emr{sou/obs}$ the component of the
source velocity along the line of sight in the observatory frame, and
$f^\emr{obs}$ the observed frequency.

Moreover, the spectrum is regularly sampled in frequency. Its frequency
axis is thus described as
\begin{equation}
  \label{eq:spectrum}
  f(i) = f_\emr{ref} + (i-i_\emr{ref})\,\delta f,
\end{equation}
where $f_\emr{ref}$ is the reference frequency at the reference channel
$i_\emr{ref}$, and $\delta f$ the frequency channel spacing. The astronomer
is interested by the description of the velocity variations in the source
rest frame. However, the spectrum is recorded in the observatory frame. The
same intensity $I(i)$ of the spectrum can thus be attributed to two
different frequencies, $f^\emr{obs}(i)$ and
$f^\emr{rest}(i)$. Equation~\ref{eq:spectrum} can thus be written in the
two frames for the same channel $i$ as
\begin{eqnarray}
  f^\emr{obs}(i)  &=& f^\emr{obs}_\emr{ref}  + (i-i_\emr{ref})\,\delta f^\emr{obs}, \\
  f^\emr{rest}(i) &=& f^\emr{rest}_\emr{ref} + (i-i_\emr{ref})\,\delta f^\emr{rest}.
\end{eqnarray}
Applying Eq.~\ref{eq:doppler:radio} yields
\begin{equation}
  f^\emr{obs}_\emr{ref} = f^\emr{rest}_\emr{ref} \, \paren{1-\frac{v_\emr{sou/obs}}{c}},
  \quad \mbox{and} \quad
  \delta f^\emr{obs} = \delta f^\emr{rest} \, \paren{1-\frac{v_\emr{sou/obs}}{c}}.
\end{equation}
On one hand, the channel spacing in the observatory frame
$(\delta f^\emr{obs})$ is fixed by the spectrometer hardware. On the other
hand, there is an infinite number of
$(i_\emr{ref},f^\emr{obs}_\emr{ref},f^\emr{rest}_\emr{ref})$ values to
describe the same spectrum. The simplest choice is to set
$f^\emr{rest}_\emr{ref}$ to the rest frequency of the line of interest,
\eg, $f^\emr{rest} = 110.20135$\,GHz for the \thCO{} \Jone{} line, and to
use $f^\emr{obs}_\emr{ref}$ as the tuning frequency of the receiver,
implying that the reference channel and thus the associated line will be
localized at the middle of the spectrum frequency axis.

The Doppler frequency shift $(f^\emr{rest}-f^\emr{obs})$ of
Eq.~\ref{eq:doppler:radio} varies with time during the day because of the
Earth rotation around its axis and during the year because of the Earth
rotation around the Sun. To remove this time dependency at the tuning
frequency, radio-observatories slightly shift the tuning frequency with
time according to the relative velocity between the observatory and the
inertial frame, named Kinematic Local Standard of Rest (LSRK). The
remaining Doppler effect between the LSRK frame and the source rest frame
is dealt with in the data reduction software because it is independent of
the observing time. However, the hardware correction, called real-time
Doppler tracking, has two main limitations.
\begin{itemize}
\item First, as it is only applied to the tuning frequency, it exactly
  corrects only the rest frequency at the reference channel while the
  radio-astronomy receivers observe wide bandwidth at high spectral
  resolution. The frequency scale in the source frame thus experiences a
  time-dependent frequency dilation around the reference frequency:
  \mbox{$\delta f^\emr{rest} = \delta
    f^\emr{obs}/\cbrace{1-v_{\text{sou/obs}}(t)/c}$}, with
  $\delta f^\emr{obs}$ the channel spacing fixed by the spectrometer
  hardware in the observatory frame. The order of magnitude of the Earth
  velocity in the LSRK frame, $|\Delta v| \le 30\kms$, implies that the
  dilation effect, $\delta f^\emr{rest}$, becomes of the order of the
  channel spacing every few tens of thousands channels. No observatory is
  yet proposing a hardware solution to correct for this dilation effect.
\item Second, when scanning the receiver over a portion of the sky to
  obtain wide-field imaging, the Doppler tracking correction is computed
  only once at the start of each scan. This is to ensure that potential
  standing wave associated with the cavity composed of, \eg, the primary
  and secondary mirrors, have a periodicity along the frequency axis that
  is fixed during the scan duration. The Doppler tracking correction is
  thus only approximate because it is computed only once every few minutes
  in a particular sky direction, while the Doppler effect continuously
  depends both on the time and sky direction. The dependence on the sky
  direction is most problematic when scanning a wide portion of sky during
  a single scan.
\end{itemize}
Correcting for the time and space dependence of the Doppler effect implies
a shift of the reference channel $(i_\emr{ref})$ at constant reference
frequency $(f^\emr{rest}_\emr{line})$ in the source frame~\citep[for
details, see, \eg,][]{pety2011}.  The observed spectra are thus slightly
shifted in frequency. Moreover, current heterodyne receivers cover
  two frequency bands located below (lower side band) and above (upper side
  band) the frequency of the local oscillator. Due to the difference in
  frequency between the two bands (16 GHz for the EMIR receiver), the
  velocity scales are slightly different for these two side
  bands. Furthermore the separation of the signals from the two bands is
  not perfect. This may lead to the apparition of “ghost” lines from the
  rejected band at frequencies that depend on the local oscillator
  frequency. We refer the reader to~\citet{pety2011} for associated
  details. All in all, the spectra thus need to be resampled to a common
frequency axis before merging them to avoid blurring the spectral response
in the science-ready product.

\section{Calibration in a nutshell}
\label{app:calibration}

In this appendix, we summarize the calibration of the raw data, which
combines the determination and application of the time varying calibration
factor with the removal of the contribution of the atmosphere to the
measured intensity. For simplicity, we start with assuming that the gain of
the measurement is constant with time before generalizing to the case where
the gain actually varies with time. We finally look at the impact of this
calibration scheme on the measured noise. We do not speak about important
additional subtleties, such as the impact of the mixing of the image
sideband into the signal sideband or the usefulness of smoothing the
frequency bandpass response when determining the calibration gain.

\subsection{Time independent gain}

The intensity measured $(I_\emr{meas})$ by the receiver can be written
before calibration and to zero order as the sum of the contribution of the
astronomical signal $(S_\emr{astro})$ and of the atmospheric emission
$(S_\emr{atm})$, multiplied by a gain $(g)$
\begin{equation}
  I^\emr{meas} = g.\paren{S^\emr{astro}+S^\emr{atm}}.
\end{equation}
The astronomer is interested to recover the astronomical signal. However,
the contribution of the atmosphere most often completely dominates the
astronomical signal at millimeter wavelengths, \ie,
\mbox{$S^\emr{astro} \ll S^\emr{atm}$}. It is thus required to measure
independently the contribution of the atmosphere in order to subtract it. A
common way to do this is to regularly observe a reference line of sight in
between the observations of the on-source lines of sight. This method is
called position switching. Writing the two observations as
\begin{eqnarray}
  \label{eq:on:1}
  \emr{ON}  &=& g.\paren{S^\emr{astro}_\emr{on} +S^\emr{atm}_\emr{on}},\\
  \label{eq:ref:1}
  \emr{REF} &=& g.\paren{S^\emr{astro}_\emr{ref}+S^\emr{atm}_\emr{ref}},
\end{eqnarray}
this gives
\begin{equation}
  \label{eq:calib:1}
  S^\emr{astro}_\emr{on} %
  = \frac{1}{g}\,\paren{\emr{ON}-\emr{REF}} %
  + S^\emr{astro}_\emr{ref} %
  + \paren{S^\emr{atm}_\emr{on} - S^\emr{atm}_\emr{ref}}. %
\end{equation}
When the reference line of sight is actually devoid of signal
\mbox{$(S^\emr{astro}_\emr{ref} = 0)$}, and the contribution from the
atmosphere is stable between the on-source and reference lines of sight,
the last two terms cancel and we obtain
\begin{equation}
  \label{eq:calib:perfect:1}
  S^\emr{astro}_\emr{on} %
  = \frac{1}{g}\,\paren{\emr{ON}-\emr{REF}}. %
\end{equation}

\subsection{Time varying gain}

This gain is a combination of the absorption of the atmosphere and of the
electronic amplification of the receiver. The electronic gain is constant
over a typical timescale of about 30 minutes. But the atmosphere absorption
varies on much shorter timescales.  Moreover the atmosphere absorption and
receiver amplification vary with frequency. In order to take into account
the time variation of the system (atmosphere + receiver) gain, we model it
as the product of the atmosphere and the receiver gain.
\begin{equation}
  g = g^\emr{rec}\,g^\emr{atm}.
\end{equation}
Using this expression in Eq.~\ref{eq:on:1} and~\ref{eq:ref:1}, we obtain
\begin{eqnarray}
  \label{eq:on:2}
  \emr{ON}  &=& g^\emr{rec}\,g^\emr{atm}_\emr{on}\,\paren{S^\emr{astro}_\emr{on} +S^\emr{atm}_\emr{on}},\\
  \label{eq:ref:2}
  \emr{REF} &=& g^\emr{rec}\,g^\emr{atm}_\emr{ref}\,\paren{S^\emr{astro}_\emr{ref}+S^\emr{atm}_\emr{ref}}.
\end{eqnarray}
In order to solve for \mbox{$S^\emr{astro}_\emr{on}$}, we first remove the
receiver dependency because it dominates the spectral part of the gain
variations, in particular at the edges of the observed bandpass. To do
this, we just take the ratio of the \emr{ON} and \emr{REF}
measurements. This yields
\begin{equation}
  \label{eq:on:over:ref}
  \frac{\emr{ON}}{\emr{REF}} %
  = \frac{g^\emr{atm}_\emr{on}}{g^\emr{atm}_\emr{ref}}\, %
  \frac{\bracket{S^\emr{astro}_\emr{on}+S^\emr{atm}_\emr{on}}}{\bracket{S^\emr{astro}_\emr{ref}+S^\emr{atm}_\emr{ref}}} %
  \sim 1.
\end{equation}
This ratio is of order 1 for two reasons.
\begin{enumerate}
\item The astronomical signal is (most often) dominated by the atmospheric
  signal, \ie, \mbox{$S^\emr{astro} \ll S^\emr{atm}$}.
\item The time variation of the gains are mostly due to variations of the
  atmosphere absorption, which are to first order anticorrelated with the
  variations of the atmosphere emission. This can be written as
  \begin{equation}
    \label{eq:anticorrelation}
    g^\emr{atm}_\emr{on}\,S^\emr{atm}_\emr{on} \sim g^\emr{atm}_\emr{ref}\,S^\emr{atm}_\emr{ref}.
  \end{equation}
  This is of course only true when the atmosphere varies only slightly
  during the observation.
\end{enumerate}
We thus subtract 1 to the ratio of Eq.~\ref{eq:on:over:ref} in order to
mimic a Taylor decomposition. Solving for \mbox{$S^\emr{astro}_\emr{on}$}
then yields the sum of three terms
\begin{equation}
  \label{eq:calib:2}
  S^\emr{astro}_\emr{on} %
  = T_\emr{sys}\,\cbrace{\frac{\emr{ON}}{\emr{REF}}-1} %
  + S^\emr{cal,astro}_\emr{ref} %
  + {\cal B},
\end{equation}
\begin{eqnarray}
  \label{eq:tsys}
  \mbox{with} &\quad& T_\emr{sys} = \frac{\emr{REF}}{g^\emr{atm}_\emr{on}},\\
  \label{eq:ref:signal}
              &\quad& S^\emr{cal,astro}_\emr{ref} =
                      \frac{g^\emr{atm}_\emr{ref}}{g^\emr{atm}_\emr{on}}\,S^\emr{astro}_\emr{ref} \sim 0,\\
  \label{eq:baseline}
  \mbox{and}  &\quad& {\cal B}
                      = S^\emr{atm}_\emr{on}\,%
                      \cbrace{%
                      \frac{g^\emr{atm}_\emr{ref}\,S^\emr{atm}_\emr{ref}}{g^\emr{atm}_\emr{on}\,S^\emr{atm}_\emr{on}} %
                      -1} %
                      \sim 0.
\end{eqnarray}
Equation~\ref{eq:calib:2} is a generalization of Eq.~\ref{eq:calib:1} to
the case where the gain varies with time during the observations. Both have
three terms.
\begin{description}
\item[\textbf{The baseline}] The term ${\cal B}$ is the residual that is
  nonzero when the assumption that the atmosphere emission and absorption
  are anticorrelated, \ie, Eq.~\ref{eq:anticorrelation}, breaks. This term
  is responsible for the typical continuum variations, called baselines,
  seen around the lines. These baseline offsets are removed through the
  baselining procedure described in Sect.~\ref{sec:noise:levels}.
\item[\textbf{The reference signal}] The term
  \mbox{$S^\emr{cal,astro}_\emr{ref}$} is exactly zero, except when there
  exists some residual signal from the astronomical source on the reference
  line of sight. This happens for lines whose emission is extended over
  several degrees on the plane of sky, for instance, the \twCO{} and
  \thCO{} \Jone{} emissions from local Giant Molecular Clouds. This is
  nevertheless rather the exception than the rule. When this term is
  nonzero, it can not be treated through baselining as the previous
  continuum offset. Indeed, it has a similar shape as the on-source
  line. It must thus be measured independently at relatively large
  signal-to-noise ratio and added back to the calibrated on-source signal.
  Contrary to common belief, this is \mbox{$S^\emr{cal,astro}_\emr{ref}$}
  that must be added, and not $S^\emr{astro}_\emr{ref}$. In other words,
  the astronomical signal towards the reference line of sight must be added
  \textit{after} multiplication by the time gain ratio between the
  on-source and reference observations.
\item[\textbf{The on-source signal}] Under perfect conditions, we recover
  an equation whose shape is similar to Eq.~\ref{eq:calib:perfect:1}, \ie,
  \begin{equation}
    \label{eq:calib:perfect:2}
    S^\emr{astro}_\emr{on} %
    = T_\emr{sys}\,\cbrace{\frac{\emr{ON}}{\emr{REF}}-1}. %
  \end{equation}
  The term in parenthesis is unitless and the system temperature
  $(T_\emr{sys})$ is the multiplicative calibration factor needed to
  establish the correct intensity unit scale. The system temperature
  depends both on frequency and time.
\end{description}

\section{Noise spatial power density}
\label{app:nspd}

The spatial energy density of a 2D stochastic process $D$ is defined as
\begin{equation}
  \ESD_D(u,v) = \Esp{\modulus{\ft{D}}^2(u,v)},
\end{equation}
where $\ft{D}$ is the Fourier transform of $D$, and $\mathbb{E}$ is the
expectation operator. In our case, the stochastic process will be the
measurement of the signal affected by random noise over an image of area
$\Area{ima}$, and the expectation will be measured as the average of the
images over a given number of channels. The spatial power density of $D$ is
then defined as the spatial energy density divided by the area of the
image, \ie,
\begin{equation}
  \PSD_D(u,v) = \frac{\ESD_D(u,v)}{\Area{ima}}.
\end{equation}

The reference spectrum is observed only in between the observation of two
consecutive lines on source.  The integration time at the reference
position is much larger than the integration time for each ON spectrum. The
contribution of the noise from the reference position to the noise of the
calibrated spectrum is thus negligible when computing the noise RMS per ON
position. However, the noise of the reference spectrum is shared by all the
ON spectra of two consecutive lines, implying a noise energy level
correlated to the scanning configuration (rectangular patterns).

Here we will first compute the first order term of the Taylor decomposition
of Eq.~\ref{eq:calib:perfect:2} at point $S^\emr{atm}$. This will then
allow us to show that the noise spatial power density is to first order the
sum of two components coming from the ON and REF noise spatial behaviors,
respectively. We will finally compute the quantitative impact of the noise
correlation introduced by the REF measurements.

\subsection{Linearisation of the measurement equation}
\label{app:linearisation}

We restart from Eq~\ref{eq:calib:perfect:2} that relates the calibrated
signal to the ON and REF measurements to show that we have to first order
for channels devoid of signal
\begin{equation}
  \label{eq:calib:perfect:3}
  S^\emr{astro}_\emr{on}(\theta_{l},\theta_{m},\theta_{l0},\theta_{m0}) %
  \simeq \bracket{B\star N_\emr{on}}(\theta_{l},\theta_{m})-\bracket{B\star N_\emr{ref}}(\theta_{l0},\theta_{m0}),
\end{equation}
where $\star$ is the convolution symbol, $B$ is the point spread function
of the telescope, and $(N_\emr{on},N_\emr{ref})$ are a couple of centered
normal random variables of same standard deviation as the atmospheric
signal on source or on reference, respectively.

To do this, we first redefine the ON and REF measurements to take into
account three things. First, we compute the noise spatial power density
only on channels devoid of line astronomical signal, \ie, we assume that
$S^\emr{astro} = 0$. Second the coupling of the telescope to the sky is
imperfect. This translates into a convolution equation. Third, the
telescope is scanning the sky when observing on source, while it always
comes back to the same position, $(\theta_{l0},\theta_{m0})$, devoid of
astronomical signal, when observing the reference. We will thus precise the
sky coordinates at which ON and REF spectra are measured. This gives
\begin{eqnarray}
  \label{eq:on:3}
  \emr{ON}(\theta_l,\theta_m)        &=& g^\emr{rec}.g^\emr{atm}_\emr{on}.\bracket{B\star S^\emr{atm}_\emr{on}}(\theta_l,\theta_m),\\
  \label{eq:ref:3}
  \emr{REF}(\theta_{l0},\theta_{m0}) &=& g^\emr{rec}.g^\emr{atm}_\emr{ref}.\bracket{B\star S^\emr{atm}_\emr{ref}}(\theta_{l0},\theta_{m0}).
\end{eqnarray}
The atmospheric signal $S^\emr{atm}$ can be considered as a normal random
variable of expectation $\Esp{S^\emr{atm}}$ and standard deviation
$\sigma^\emr{atm}$. As explained above
\mbox{$\sigma^\emr{atm}/\Esp{S^\emr{atm}} \ll 1$} in millimeter
radio-astronomy.  In order to prepare to compute the Taylor decomposition
of Eq.~\ref{eq:calib:perfect:3} in $\Esp{S^\emr{atm}}$, we rewrite the
atmospheric random variable as
\begin{equation}
  \label{eq:taylor:definitions}
  S^\emr{atm} = \Esp{S^\emr{atm}}\,\paren{1+\Delta},
  \quad \mbox{with} \quad
  \Delta = N/\Esp{S^\emr{atm}},
\end{equation}
where $\Delta$ is a centered normal random variable of standard deviation
$\ll 1$. Replacing the definitions~\ref{eq:taylor:definitions} in
Eq.~\ref{eq:on:3} and~\ref{eq:ref:3}, and using the fact that the integral
of $B$ is equal to one, we yield
\begin{equation*}
  \frac{\emr{ON}(\theta_l,\theta_m)}{\emr{REF}(\theta_{l0},\theta_{m0})} =
  \frac{g^\emr{atm}_\emr{on}.\Esp{S^\emr{atm}_\emr{on}(\theta_l,\theta_m)}}{g^\emr{atm}_\emr{ref}.\Esp{S^\emr{atm}_\emr{ref}(\theta_{l0},\theta_{m0})}}\,
  \cbrace{\frac{1+\bracket{B\star \Delta_\emr{on}}(\theta_l,\theta_m)}{1+\bracket{B\star\Delta_\emr{ref}}(\theta_{l0},\theta_{m0})}}.
\end{equation*}
Using again the fact that the first term of the product is of order 1, and
keeping only the first order term in the Taylor decomposition of the second
product term, we find
\begin{equation}
  \label{eq:calib:perfect:4}
  \frac{\emr{ON}(\theta_l,\theta_m)}{\emr{REF}(\theta_{l0},\theta_{m0})}
  - 1 \simeq \bracket{B\star\Delta_\emr{on}}(\theta_{l},\theta_{m})-\bracket{B\star\Delta_\emr{ref}}(\theta_{l0},\theta_{m0}).
\end{equation}
We obtain Eq.~\ref{eq:calib:perfect:3} by 1) replacing this equation in
Eq.~\ref{eq:calib:perfect:2}, 2) using the definition of $N$ in
Eq.~\ref{eq:taylor:definitions}, and 3) recognizing that
\mbox{$T_\emr{sys} \sim \Esp{S^\emr{atm}}$}.

\subsection{Normalization of the pixel variances}
\label{app:noise:norm}

We plan to compute the spatial power density of $S^\emr{astro}_\emr{on}$.
Equation~\ref{eq:calib:perfect:3} indicates that the measured signal is to
first order the subtraction of two central normal random variables of
standard deviation $\sigma_\emr{on}$ and $\sigma_\emr{ref}$. The standard
deviation of $S^\emr{astro}_\emr{on}$ is thus
\begin{equation}
  \sigma = \sqrt{\sigma_\emr{on}^2+\sigma_\emr{ref}^2}.
\end{equation}
As the weather and the source elevation is varying during the observations,
$\sigma$ varies with times and thus with the position in the final map as
shown on Fig.~\ref{fig:data:noise:dist}. Fortunately, the observing
conditions can be considered constant between the observation on source and
on reference. This implies that
\begin{equation}
  \sigma_\emr{ref} = \sigma_\emr{on}\,
  \sqrt{\frac{dt_\emr{on}}{dt_\emr{ref}}}
  \quad \mbox{and} \quad
  \sigma = \sigma_\emr{on} \, \sqrt{1+\frac{dt_\emr{on}}{dt_\emr{ref}}}
\end{equation}
where $dt_\emr{on}$ and $dt_\emr{ref}$ are the integration time on source
and on reference. We can thus compute the spatial power density of the
ratio $S^\emr{astro}_\emr{on}/\sigma$ to get rid of the noise variations
due to the weather or the source elevation. This simplifies the
interpretation of the result.

From this point on, we will keep the notation of
Eq.~\ref{eq:calib:perfect:3}, and just note that the random variables
$N_\emr{on}$ and $N_\emr{ref}$ have for standard deviation
\mbox{$\sigma_\emr{on}/\sigma$ and $\sigma_\emr{ref}/\sigma$}.

\subsection{Relative contributions from the ON and REF measurements}
\label{app:noise:on:vs:ref}

The spatial power density of the difference of two independent random
processes is the sum of the spatial power density of each random
process. We thus get
\begin{equation}
  \PSD_{S^\emr{astro}_\emr{on}} \simeq \PSD_\emr{on} + \PSD_\emr{ref},
\end{equation}
with
\begin{equation}
  \PSD_\emr{on}(u,v)
  = \frac{\Esp{\modulus{\ft{B\star N_\emr{on}}}^2(u,v)}}{\Area{ima}}
  = \Area{pix}\,\paren{\frac{\sigma_\emr{on}}{\sigma}}^2\,\modulus{\ft{B}}^2(u,v),
\end{equation}
and
\begin{equation}
  \PSD_\emr{ref}(u,v)
  = \frac{\Esp{\modulus{\ft{B\star N_\emr{ref}}}^2(u,v)}}{\Area{ima}}.
\end{equation}
The noise spatial power density of the on-source noise delivers the usual
result, \ie, it is proportional to the Fourier transform of the point
spread function of the telescope. In the next section, we will compute the
noise spatial power density of the reference noise. In particular, it is
not proportional to $\modulus{\ft{B}}^2(u,v)$ because the reference
position is always observed at the same position on sky.

\subsection{Quantitative impact of the noise correlation}
\label{app:noise:correlations}

\FigSchemaSpatialNoiseComputing{} %

Let's assume that all on-source pixels inside a rectangle of area
$\Area{rect}$ share the same reference spectrum, and that these rectangles
form a chessboard pattern. The images are paved by $\numb{rect}$ such
rectangles. The impact ($R$) of the reference observations on the
observation procedure can be modeled by a convolution of a random dirac
comb $\Sha$ with a 2D rectangular shape $\Pi$
\begin{equation}
  R(\theta_l,\theta_m) = [\Pi\star\Sha](\theta_l,\theta_m),
\end{equation}
\begin{equation}
  \label{eq:sha}
  \mbox{with} \quad %
  \Sha(\theta_l,\theta_m) = \Area{rect}\,\sum_{k=1}^{\numb{rect}} N_{k}\;\delta(\theta_l-\theta_{l,k},\theta_m-\theta_{m,k}),
\end{equation}
and
\begin{equation}
  \label{eq:pi}
  \Pi(\theta_l,\theta_m) = \left\{\begin{array}{cl} 1/\Area{rect} & \text{if } \modulus{\theta_l}<\frac{\Delta\theta_l}{2}, \text{ and } \modulus{\theta_m}<\frac{\Delta\theta_m}{2},\\ 0 & \text{else}, \end{array}\right.
\end{equation}
where $k$ is the index of one rectangle over the chessboard,
$(\theta_{l,k},\theta_{m,k})$ the position of center of each rectangle, and
$N_{k}$ is the random variable associated with each measurement of the
reference position. This random variable is assumed to be a normal variable
${\cal N}(0,\sigma_{k}^\emr{ref})$. The factor $1/\Area{rect}$ that appears
in the definition~\ref{eq:pi} ensures that the integral of the 2D
rectangular shape $\Pi$ is equal to 1 (unitless), and that all the energy
of the stochastic process $R$ is contained into the energy of the random
comb function $\Sha$. Figure~\ref{fig:schema:spatial:noise} illustrates
this decomposition in the 1D case. We will now show that the noise spatial
power density of the process $R$ is
\begin{equation}
  \label{eq:psd:R}
  \PSD_{R}(u,v) = \left[\sinC{\frac{\Delta\theta_l\,u}{\lambda}}\,\sinC{\frac{\Delta\theta_m\,v}{\lambda}} \right]^2\,\Area{rect}\,\langle\sigma_{k}^2\rangle,
\end{equation}
where
\begin{equation}
  \label{eq:mean:sigma}
  \langle\sigma_{k}^2\rangle = \frac{1}{\numb{rect}}\,\sum_{k=1}^{\numb{rect}} \sigma_k^2
\end{equation}
is the average of the noise variances of the reference measurements.

Indeed, the properties of the Fourier transform and the deterministic
nature of the $\Pi$ function allow us to yield
\begin{equation}
  \PSD_{R}(u,v) = \left|\ft{\Pi}\right|^2(u,v)\cdot\PSD_{\Sha}(u,v).
\end{equation}
As the Fourier transform of a boxcar is a cardinal sine,
\begin{equation}
  \label{eq:tf:pi}
  \left|\ft{\Pi}\right|^2(u,v) = \bracket{%
    \sinC{\frac{\Delta\theta_l\,u}{\lambda}}\,\sinC{\frac{\Delta\theta_m\,v}{\lambda}}
  }^2.
\end{equation}
As $\Sha$ is a white noise, its energy spectral density is a constant. By
using the energy conservation property of the Fourier transform, we get
\begin{equation}
  \label{eq:psd:sha}
  \PSD_{\Sha}(u,v) = \frac{\Area{rect}^2}{\Area{Ima}}\,\sum_{k=1}^{\numb{rect}}\Esp{N_k^2}.
\end{equation}
Finally, using Eq.~\ref{eq:mean:sigma} and the fact that
\mbox{$\Area{ima} = \Area{rect}\,\numb{rect}$}, the spatial power
distribution of the 2D random comb is
\begin{equation}
  \PSD_{\Sha}(u,v) = \numb{rect}\,\frac{\Area{rect}^2}{\Area{ima}}\,\langle\sigma_{k}^2\rangle
  = \Area{rect}\,\langle\sigma_{k}^2\rangle.
\end{equation}


\section{Calibration uncertainty}
\label{app:calibration:uncertainty}

\FigCalibrationUncertainty{} %

In order to monitor the calibration uncertainty, we observed the same
position with known and bright line intensities at the start of each 8-hour
block of observations. We choose the Horsehead core position (located at
$(+20'',+22'')$ from the projection center of
\radec{05}{40}{54.270}{-02}{28}{00.00}) as this position has been
extensively studied in the framework of the Horsehead WHISPER
survey~\citep[see, e.g.,][]{guzman12,pety12,gratier13,guzman13}. We used
the symmetric position switching observing mode with a reference position
located at $(-100'',0'')$ from the projection center. We integrated 6
minutes in total (3 minutes on source and 3 minutes on reference). This
yields 55 measurements times two polarization spread over slightly more
than 6 years and observed during varying weather conditions.

Calibration and reduction were done using standard methods of MRTCAL and
CLASS. After extracting 11\,MHz around the \thCO{} \Jone{} rest frequency,
we averaged the 110 separate spectra. The mean spectrum of all 110
measurements was fitted with a single Gaussian to get a reference value for
the \thCO{} \Jone{} line integrated intensity. In this fit, we only
considered the main component of Orion B, near $10\kms$. We then fitted a
single Gaussian for each 6-minutes measurement using the solution for the
averaged spectra as initial guess for the fit and we visually checked that
all fits were good. We then computed the variation of each measurement
relative to the value derived for the average
spectrum. Figure~\ref{fig:data:calibration:uncertainty} shows the relative
variations as a function of the time and their histograms for the H and V
EMIR polarizations separately.

The relative variations range from $-25$ to $+10$\%. The vertical
polarization delivers almost systematically a larger intensity than the
horizontal polarization. This explains why the mean relative variations are
$+1.01$ and $-1.09$\%, respectively. The RMS around these means
are 6.8 and 6.1\%. The median absolute deviations are 5.0 and 4.3\%,
respectively. The difference between the RMS and median absolute deviation
implies that a few measurements are outliers. Overall, the calibration
uncertainty for the IRAM 30m is of the order of 5\% for an almost
instantaneous observation at 3\,mm, as is the case for the ORION-B
  project. Averaging many such measurements when, \textit{e.g.},
  observing low brightness sources, considerably reduces this
uncertainty. This experiment says nothing about the absolute calibration
accuracy.


\section{About the multilayer perceptron}
\label{app:perceptron}

\FigPerceptron{}%

A neural network is a graphical model that maps nonlinearly outputs from
inputs. A neural network architecture represents a given class of function.
Neural networks are composed of a succession of layers that perform
nonlinear transformations.  Figure~\ref{fig:perceptron} shows an example
of a forward propagation architecture, named multilayer perceptron. Forward
propagation means that the output of one layer is always the input of the
next layer, while multilayer implies that there are at least two layers. In
addition, the layers are ``fully connected'', \ie{} each output of a layer
is connected by a linear combination of the inputs. More precisely, for
each layer, the input $x$ and the output $y$ are related through
\begin{equation}
  y = \sigma\left(Wx + b\right),
\end{equation}
where $W$ and $b$ are respectively the weight matrix and the bias vector,
which are estimated during the learning step. The biases enable to shift
the argument of a nonlinear activation function $\sigma$.  Such a
perceptron has the universal approximation property, \ie{}, it is able to
approximate as precisely as required any continuous function provided it
has enough neurons.


\section{A short introduction to mutual information}
\label{app:mutual:information}

In information theory, mutual information is a quantity that measures the
statistical dependence of two variables. In the case of continuous real
variables, the mutual information in base 2 is calculated as follows:
\begin{equation}
  \mi(X,Y) = \int\displaylimits_{-\infty}^{+\infty} \int\displaylimits_{-\infty}^{+\infty} f_{X,Y}(x,y)\,\log \frac{f_{X,Y}(x,y)}{f_X(x)f_Y(y)}.
\end{equation}
where $f_X$ and $f_Y$ are respectively the probability density functions of
$X$ and $Y$, and $f_{X,Y}$ the probability density function of the couple
$(X,Y)$.

Mutual information is a positive real quantity and it is symmetric, \ie,
$\mi(X,Y) = \mi(Y,X)$. If there exists a function $f$ (linear or not) such
that $Y = f(X)$, then $\mi(X,Y) = +\infty$. Conversely, if the knowledge of
one of the variables gives no information about the other (\ie, the two
variables are independent) then $\mi(X,Y) = 0$.  Mutual information is
therefore a more general indicator than Pearson or Spearman correlation
coefficient since it takes into account nonlinear and nonmonotonic
relationships. In particular, it is possible for two variables to be
decorrelated but have nonzero mutual information.


\section{Supplementary figures}

\FigAllChannelsRadio{} %


\end{appendix} %

\end{document}